\documentclass[11pt,a4paper]{article}
\usepackage{jheppub,epstopdf,bbm,slashed,tensor}

\usepackage{simplewick} 
\usepackage{amsmath,amssymb,epsfig,graphicx}

\makeatletter
\newcommand{\fmslash}[2][0mu]{
\mathchoice
    {\fmsl@sh\displaystyle{#1}{#2}}%
    {\fmsl@sh\textstyle{#1}{#2}}%
    {\fmsl@sh\scriptstyle{#1}{#2}}%
   
    {\fmsl@sh\scriptscriptstyle{#1}{#2}}}
\newcommand{\fmsl@sh}[3]{%
\m@th\ooalign{$\hfil#1\mkern#2/\hfil$\crcr$#1#3$}}

\newcommand{\tr}{\hbox{tr}}

\title{\center{Quantum noncommutative ABJM theory: first steps}} 

\author[a,1]{Carmelo P. Martin,
\note{carmelop@fis.ucm.es}}
\affiliation[a]{Departamento de F\'{\i}sica Te\'{o}rica I, Facultad de Ciencias F\'{\i}sicas,
Universidad Complutense de Madrid, 28040-Madrid, Spain}
\author[b,c,2]{Josip Trampetic
\note{josip.trampetc@irb.hr}}
\affiliation[b]{Rudjer Bo\v skovi\' c Institute, Division of Experimental Physics, HR-10002 Zagreb, Croatia}
\affiliation[c]{Max-Planck-Institut f\"ur Physik, (Werner-Heisenberg-Institut), F\"ohringer Ring 6, D-80805 M\"unchen, Germany}
\author[d,3]{ and  Jiangyang You
\note{jiangyang.you@irb.hr}}
\affiliation[d]{Rudjer Bo\v skovi\' c Institute, Division of Physical Chemistry, HR-10002 Zagreb, Croatia}

\abstract{We introduce ABJM  quantum field theory in the noncommutative spacetime by using the component formalism and show that it is ${\cal N}=6$ supersymmetric. For the $\rm U(1)_\kappa\times U(1)_{-\kappa}$ case, we compute all one-loop 1PI two- and three-point functions in the Landau gauge and show that they are UV finite and have  well-defined commutative limits $\theta^{\mu\nu}\rightarrow 0$, corresponding exactly to the 1PI functions of the ordinary ABJM field theory. This result also holds for all one-loop functions which are UV finite by power counting. It seems that the noncommutative quantum  ABJM field theory is free from the noncommutative IR instabilities.}

\keywords{Non-Commutative Geometry, Supersymmetry, Chern-Simons theory }

\begin{document}

\maketitle

\section{Introduction}

ABJM field theory at the level $\kappa$ was introduced in \cite{Aharony:2008ug} to provide a holographic dual of the M theory on the $AdS_4\times S^7/Z_k$, thus furnishing a concrete realization of the famous gauge/gravity duality conjecture \cite{Maldacena:1997re}. From the point of view of Quantum Gravity, ABJM quantum field theory deserves being analyzed thoroughly since it affords  possibility of studying gravity on four dimensional spacetime at the quantum level \cite{Drukker:2010nc,Dabholkar:2014wpa}. Besides, the ABJM theory may be useful in the effective field theory description of a certain condensed matter systems where the Chern-Simons action arises naturally \cite{Bea:2014yda, Ammon:2015wua}.

That the ABJM field theory, at the level $\kappa$, carries the ${\cal N} =6$ super-conformal symmetry was explicitly verified in the component formalism formulation of the theory \cite{Bandres:2008ry}.  This super-conformal symmetry  is enhanced to ${\cal N} =8$, when  $\kappa=1$ or $\kappa=2$ \cite{Kwon:2009ar}.  In Ref. \cite{Buchbinder:2008vi} ABJM theory was formulated in the ${\cal N} =3$ harmonic superspace. Such formulation was used to show that the super-field perturbation theory, obtained in the background field formalism for the background field gauge,  is UV finite \cite{Buchbinder:2009dc}.

Modern on-shell techniques have been used to work out some tree-level \cite{Bargheer:2010hn}  and one-loop \cite{Bargheer:2012cp,Bianchi:2012cq} scattering amplitudes in the ABJM field theory. These computations have unveiled beautiful algebraic --the Yangian of the corresponding super-conformal algebra \cite{Bargheer:2010hn, Lee:2010du}-- and geometric --the orthogonal Grassmannian \cite{Elvang:2014fja}-- structures that  play an important role to the analysis of the theory: its integrability in particular  \cite{Beisert:2017pnr}.

Noncommutative field theory --see \cite{Szabo:2001kg}, for a review-- is a well-established area of research in High Energy Theoretical Physics. Surprisingly, to the best of our knowledge, no formulation of the ABJM field theory on noncommutative spacetime  can be found in the existing literature. This state of affairs should not continue, since quantum ABJM field theory on the noncommutative spacetime could be helpful --through the gauge/gravity correspondence-- in studying noncommutative gravity in four  dimensions and, on the other hand, noncommutative Chern-Simons theory  naturally arises in the study of the Fractional Quantum Hall effect \cite{Fradkin:2002qw}. Further, noncommutative spacetime as defined by the Moyal product breaks conformal invariance, so that one may discuss in a well-defined setting wether or not the beautiful structures and effects that occur in the ABJM field theory disappear together with the disappearances of the super-conformal invariance of it or, perhaps, are replaced by some noncommutative ones. 

Main purpose of this paper is to formulate the complete ABJM quantum field theory on the noncommutative spacetime as defined by the so-called Moyal star-product, via star commutator of the coordinates $[x^\mu\stackrel{\star}{,}x^\nu]=i\theta^{\mu\nu}$, with $\theta^{\mu\nu}$ being the noncommutativity matrix. We shall do this in the component formalism and show that both the classical ABJM action and the ${\cal N} =6$ ordinary transformations \cite{Bandres:2008ry} can be nicely generalized to the Moyal noncommutative spacetime to define a noncommutative ABJM quantum field theory with ${\cal N} =6$ supersymmetry.

As already mentioned, ABJM theories are proposed as the holographic dual of M2 brane in appropriate backgrounds. The noncommutative deformation of the gravity dual of the ordinary ABJM theory was worked out in \cite{Imeroni:2008cr}. Recently, it has also been shown in \cite{Colgain:2016gdj} that this B-field charged IIA supergravity background, for generic $\kappa$'s, poses the same amount of supersymmetry as its ordinary ${\cal N}=6$ counterpart does. Therefore, as will be shown below, by possessing six supersymmetries our noncommutative ABJM (NCABJM) action does fulfill the necessary condition to become dual to the superstring/supergravity theory on the deformed background constructed in \cite{Imeroni:2008cr}.

Another important aim of this paper is to check on the quantum level, whether the limit $\theta^{\mu\nu}\to 0$ of the noncommutative ABJM theory restores back the ordinary/commutative ABJM theory introduced in \cite{Aharony:2008ug}. We shall do this by computing all one-loop 1PI functions involving fewer than four fields in the noncommutative variant of  the $\rm U(1)_{\kappa}\times U(1)_{-\kappa}$ theory. This is a nontrivial issue for the following reasons:   In the component formalism the 1PI Green functions are not UV finite by power counting and, therefore, one cannot use Lebesgue's dominated convergence theorem to take limit $\theta^{\mu\nu}\to 0$ inside the integral. Actually, the   expected UV finiteness of the theory comes as a result of  cancellations that occur upon summing over all the planar parts of  the UV divergent Feynman diagrams contributing to a given 1PI Green function. Now, due to the UV/IR mixing the nonplanar part of each UV divergent Feynman diagram contributing to a given 1PI function develops, in general, a noncommutative IR divergence; only upon adding up all those noncommutative IR divergent contributions one may expect that the final  noncommutative IR divergence goes away completely. Of course, when cancellation of infinities takes place by summing up all contributions, local finite parts of the 1PI Green functions may not be uniquely defined. What is more, Moyal phases act as UV regulators of the nonplanar contributions --trading an UV divergence for an IR one-- but they are regulators which break Lorentz invariance,  so that structures of the finite contributions arising from them are not given by the standard results in renormalization theory. Actually, values of some integrals contributing to a certain Feynman diagram --see appendix C.2, for example-- remains bounded as one approaches  $\theta^{\mu\nu}=0$ point, but the $\theta^{\mu\nu}\to 0$ limit does not exist.  Putting it all together, we conclude that it is far from clear that the limit $\theta^{\mu\nu}\to 0$ of the 1PI Green functions in the noncommutative formulation of the ABJM quantum field theory are the corresponding functions in the commutative ABJM quantum field theory.

Layout of this paper is as follows: In section 2, we describe the field contents of the ordinary/classical $\rm U(1)_{\kappa}\times U(1)_{-\kappa}$ ABJM field theory action to set the notation and conventions regarding the  global $\rm SU(4)$ R-symmetry of the ABJM  theory without the notation complications due to the use of the $\rm U(N)$ groups.  Classical action of noncommutative $\rm U(N)_{\kappa}\times U(N)_{-\kappa}$ ABJM field theory is introduced next, along with the noncommutative BRST transformations which leave that action invariant --subsections 2.1 and 2.2. Noncommutative ${\cal N} =6$ supersymmetry transformations which leave the classical action of the ABJM theory invariant are introduced in subsection 2.3. In the appendix A we display a detailed proof that supersymmetric  transformations introduced in subsection 2.3 do indeed leave the classical noncommutative ABJM action invariant. Also in the appendix A we consider only the $\rm U(1)_{\kappa}\times U(1)_{-\kappa}$ case since the generalization to the $\rm U(N)_{\kappa}\times U(N)_{-\kappa}$ is straightforward and, besides, it is for the  $\rm U(1)_{\kappa}\times U(1)_{-\kappa}$ case that the difference between the classical action of the noncommutative ABJM theory and the ordinary ABJM theory is more conspicuous, due to the fact that the Moyal star-product is not commutative and generates nonabelian gauge symmetry. Feynman rules for the noncommutative $\rm U(1)_{\kappa}\times U(1)_{-\kappa}$ ABJM quantum field theory in Landau gauge are given in section 3. Power counting rules and limit $\theta^{\mu\nu}\to 0$ were discussed in section 4, while remaining rules relevant to our computations are given in the appendix D. Let us point out that we quantize the theory in the Landau gauge for two reasons: {\it i)} the Chern-Simons propagator is simpler and {\it ii)} it does not contain contributions with a dangerous IR behaviour --see section III of Ref. \cite{Pisarski:1985yj}. In sections 5 to 12 we show and discuss that, at the one-loop level all the 1PI two and three point functions of the noncommutative $\rm U(1)_{\kappa}\times U(1)_{-\kappa}$ ABJM quantum field theory are UV finite and have  well-defined limits when $\theta^{\mu\nu}\to 0$, and that those limits are equal to the corresponding Green functions of the commutative ABJM quantum field theory.  Remaining appendices are needed for properly understanding the main text.

\section{Classical NCABJM field theory}

We begin our construction for NCABJM field theory from its field contents, which is identical to the commutative theory, although the fields are noncommutative. For this reason and the convenience of comparison we briefly summarize the known results on the commutative ABJM theory first. Our conventions follow exactly those in~\cite{Bandres:2008ry}. We start with the $\rm U(1)_{\kappa}\times U(1)_{-\kappa}$ theory since it has less indices and thus it is simpler with respect to the general $\rm U(N)_{\kappa}\times U(N)_{-\kappa}$ field theory.

The pair of the $\rm U(1)_{\kappa}\times U(1)_{-\kappa}$ vector gauge fields are denoted as $A_\mu$ and $\hat A_\mu$, i.e. gauge and hgauge fields, respectively. Scalars $X_A$ and fermions $\Psi^A$ have $\rm U(1)$  charges $(+,-)$, while their adjoints have charges $(-,+)$, respectively. As in constructing the full $\rm U(N)_{\kappa}\times U(N)_{-\kappa}$ theory with above convention we choose to normalize fields so that the $\kappa$-level Lagrangian is $\kappa$ times the level-1 Lagrangian. Thus the N=1 action is as given below:
\begin{equation}
S = \frac{\kappa}{2\pi} \int d^3 x \Big( - D^\mu X^A D_\mu X_A
+i\bar\Psi_A \slashed{D} \Psi^A + \frac{1}{2}\epsilon^{\mu\nu\lambda} \big(A_\mu \partial_\nu A_\lambda
- \hat A_\mu \partial_\nu \hat A_\lambda\big) \Big),
\label{3}
\end{equation}
with four complex scalars $X_A$ and their adjoints $X^A$, where a lower index labels the
${\bf 4}$ representation and an upper index labels the complex-conjugate ${\bf \bar 4}$ representation of the global $\rm SU(4)$ R-symmetry, respectively.
Covariant derivative acting on scalar fields $X_A$  and $X^A$ respectively reads:
\begin{equation}
D_\mu X_A^{(A)} = \partial_\mu X_A^{(A)} {\stackrel{(-)}{+}} i (A_\mu - \hat A_\mu) X_A^{(A)}.
\label{2}
\end{equation}
The above pair of two-component fermi fields with notation $\bar\Psi^A$ or $\bar\Psi_A$, in (\ref{3}), implies transposing the spinor index of $\Psi^A$ and $\Psi_A$, respectively and right multiplication by $\gamma^0$ respectively, though that index is not displayed. In this definition there is no additional complex conjugation, since the lower index indicates the ${\bf 4}$ and an upper index indicates the ${\bf \bar 4}$ representation, respectively. With these conventions identities that hold for Majorana spinors shall be used for our spinors, as well, even though they are Dirac-complex fields: $\bar\Psi^A \Psi_B = \bar\Psi_B \Psi^A$. Considering Pauli-Dirac algebra conventions our $2\times 2$ Dirac
matrices satisfy $\{\gamma^\mu,\gamma^\nu\} = 2\eta^{\mu\nu}$. Here index $\mu=0,1,2$ is 3-dimensional Lorentz index with signature $(-,+,+)$. Using a Majorana representation implies that $\gamma^\mu$ is real, while choices $\gamma^0 = i\sigma^2$, $\gamma^1 =\sigma^1$, $\gamma^2 =\sigma^3$ and $\gamma^{\mu\nu\lambda} = \epsilon^{\mu\nu\lambda}$, gives $\gamma^0 \gamma^1 \gamma^2 =1$.

General $\rm U(N)_{\kappa}\times U(N)_{-\kappa}$ ABJM theory consists of four $\rm N \times N$ matrices of complex scalars $(X_A)^a{}_{\dot{a}}$ and their adjoints $(X^A)^{\dot{a}}{}_a$, as well as the spinor field matrices $(\Psi^A)^a{}_{\dot{a}}$ and their adjoints $(\Psi_A)^{\dot{a}}{}_a$, respectivly.  They both transform as $({\bf \bar N}, {\bf N})$ and $({\bf N}, {\bf \bar N})$ representations of the gauge group, respectively.  Pair of the $\rm U(N)$ gauge fields are hermitian matrices $(A_\mu)^a{}_b$
and $(\hat A_\mu)^{\dot{a}}{}_{\dot{b}}$, respectively. In matrix notation, the covariant derivatives for scalars are
\begin{equation}
D_\mu X_A^{(A)} = \partial_\mu X_A^{(A)} {\stackrel{(-)}{+}} i (A_\mu X_A^{(A)} - X_A^{(A)}\hat A_\mu),
\label{5}
\end{equation}
while for spinor fields we have equivalent expressions. Infinitesimal gauge transformations are given by
\begin{equation}
\delta A_\mu = D_\mu \Lambda = \partial_\mu \Lambda+i [A_\mu,\Lambda],\;
\delta \hat A_\mu = D_\mu \hat\Lambda = \partial_\mu\hat\Lambda +i [\hat A_\mu,\hat\Lambda],\;
\delta X_A =-i\Lambda X_A +i X_A \hat\Lambda,
\label{8}
\end{equation}
and so forth. For the general action see the subsections below as well as~\cite{Bandres:2008ry,Kwon:2009ar}.

\subsection{Noncommutative BRST transformations}

We now move on to the noncommutative theory by specifying its gauge symmetry in the BRS convention. Let us first introduce space spanned by the Moyal star($\star$)-product
\begin{equation}
 (f\star g)(x)=f(x)\star g(x)= f(x) e^{\frac{i}{2}
 \stackrel{\leftarrow}{\partial_\mu}\theta^{\mu\nu}\stackrel{\rightarrow}{\partial_\nu}} g(x),
\label{Mstar}
\end{equation}
and the following multiplication consistency relations, 
\begin{equation}
X_A\star X^B \longrightarrow (X_A)^{a}{}_{\dot{b}}
\star (X^B)^{\dot{b}}{}_{b}\;,\;{\rm and}\;\;
X^B\star X_A \longrightarrow (X^B)^{\dot{a}}{}_{a}
\star (X_A)^{a}{}_{\dot{b}},
\label{Multcon}
\end{equation}
hence the Moyal star-product of four $X$'s reads as
\begin{equation}
X_A\star X^B \star X_C\star X^D,\;\;{\rm and}\;\;
X^A\star X_B \star X^C\star X_D.
\label{Multprod}
\end{equation}
It is also worth noting that the maximum (nondegenerate) rank of the matrix, $\theta^{\mu\nu}$, is 2, since we are in three dimensions.  To avoid unitarity problems --see \cite{Gomis:2000zz,Aharony:2000gz},  we shall assume that $\theta^{0i}=0$, i.e., the time-space coordinate commutes. This assumption in three dimensions constrains nontrivial components of $\theta^{\mu\nu}$ to $\theta^{12}(\not=0)$ component only.

Now we define all noncommutative BRST transformations we need in the rest of this article:
\begin{equation}
\begin{array}{l}
(sA_\mu)^{a}{}_b=(D_\mu\Lambda)^{a}{}_b=(\partial_\mu\Lambda)^{a}{}_b+i[A_\mu\stackrel{\star}{,}\Lambda]^{a}{}_b,\\[8pt]
(s\hat A_\mu)^{\dot{a}}{}_{\dot{b}}=(D_\mu\hat\Lambda)^{\dot{a}}{}_{\dot{b}}=(\partial_\mu\hat\Lambda)^{\dot{a}}{}_{\dot{b}}+i[\hat A_\mu\stackrel{\star}{,}\hat\Lambda]^{\dot{a}}{}_{\dot{b}},\\[8pt]
(sX_A)^{a}{}_{\dot{a}}=-i\Lambda^{a}{}_b\star(X_A)^{b}{}_{\dot{a}}
+i(X_A)^{a}{}_{\dot{b}}\star\tilde\Lambda^{\dot{b}}{}_{\dot{a}},\\[8pt]
(s X^A)^{\dot{a}}{}_a=i(X^A)^{\dot{a}}{}_b\star\Lambda^{b}{}_a
-i\tilde\Lambda^{\dot{a}}{}_{\dot{b}}\star(X^A)^{\dot{b}}{}_a,\\[8pt]
(s\Psi^A)^{a}{}_{\dot{a}}=-i\Lambda^{a}{}_b\star(\Psi^A)^{b}{}_{\dot{a}}
+i(\Psi^A)^{a}{}_{\dot{b}}\star\tilde\Lambda^{\dot{b}}{}_{\dot{a}},\\[8pt]
(s\Psi_A)^{\dot{a}}{}_a=i(\Psi_A)^{\dot{a}}{}_b\star\Lambda^{b}{}_b
-i\tilde\Lambda^{\dot{a}}{}_{\dot{b}}\star(\Psi_A)^{\dot{b}}{}_a,\\[8pt]
s\Lambda=-i\Lambda\star\Lambda,\;\;
s\hat\Lambda=-i\hat\Lambda\star\hat\Lambda,
\end{array}
\label{BRST}
\end{equation}
with covariant derivatives being as follows
\begin{equation}
\begin{array}{l}
(D_\mu X_A)^{a}{}_{\dot{a}}=\partial_\mu(X_A)^{a}{}_{\dot{a}}+i(A_\mu)^{a}{}_b\star(X_A)^{b}{}_{\dot{a}}
-i(X_A)^{a}{}_{\dot{b}}\star(\hat A_\mu)^{\dot{b}}{}_{\dot{a}}\,,\\[8pt]
(D_\mu X^A)^{\dot{a}}{}_a=\partial_\mu(X^A)^{\dot{a}}{}_a +i(\hat A_\mu)^{\dot{a}}{}_b\star (X^A)^{b}{}_a
-i(X^A)^{\dot{a}}{}_{\dot{b}}\star(A_\mu)^{\dot b}{}_a\,,\\[8pt]
(D_\mu\Psi^A)^{a}{}_{\dot{a}}=\partial_\mu(\Psi^A)^{a}{}_{\dot{a}}+i(A_\mu)^{a}{}_b\star(\Psi^A)^{b}{}_{\dot{a}}
-i(\Psi^A)^{a}{}_{\dot{b}}\star(\hat A_\mu)^{\dot{b}}{}_{\dot{a}}\,,\\[8pt]
(D_\mu\Psi_A)^{\dot{a}}{}_a=\partial_\mu(\Psi_A)^{\dot{a}}{}_a+i(\hat A_\mu)^{\dot{a}}{}_b\star(\Psi_A)^{b}{}_a
-i(\Psi_A)^{\dot{a}}{}_{\dot{b}}\star(A_\mu)^{\dot{b}}{}_a\;.
\end{array}
\label{CovD}
\end{equation}

\subsection{Noncommutative generalization of the action}

Our next step is to present the classical action of NCABJM field theory. From now on we restrict ourselves to $\rm {U(1)_{\kappa} \times U(1)_{-\kappa}}$ theory for simplicity, since, generalization to $\rm {U(N)_{\kappa} \times U(N)_{-\kappa}}$ is straightforward because of the multiplication consistency relations \eqref{Multcon}. This action consists of terms that are  generalizations of those of ordinary $\rm {U(1)_{\kappa} \times U(1)_{-\kappa}}$ ABJM field theory, as well as the new interaction terms that are analogous to the commutative $\rm {U(N)_{\kappa} \times U(N)_{-\kappa}}$ theory yet vanish for N=1.  The noncommutative Chern--Simons, kinetic and additional terms having four and six fields respectively, are
\begin{eqnarray}
S&=&S_{\rm CS}+S_{\rm kin}+S_{4}+S_{6},
\label{Action}\\
S_{\rm CS} &=& \frac{\kappa}{2\pi} \int d^3 x  \, \epsilon^{\mu\nu\lambda} \tr\bigg(
\frac{1}{2} A_\mu\star \partial_\nu A_\lambda  + \frac{i}{3} A_\mu\star A_\nu\star A_\lambda-\frac{1}{2} \hat A_\mu\star \partial_\nu \hat A_\lambda - \frac{i}{3} \hat A_\mu\star \hat A_\nu\star \hat A_\lambda\bigg),
\nonumber\\
\label{ACS}\\
S_{\rm kin} &=& \frac{\kappa}{2\pi} \int d^3 x \,\tr\left( - D^\mu X^A\star D_\mu
X_A +i\bar\Psi_A \star\slashed{D} \Psi^A\right),
\label{Akin}\\
S_{4}&=&S_{4a}+S_{4b}+S_{4c},
\label{AS4}\\
S_{4a}&=&\frac{i\kappa}{2\pi} \int d^3 x\,\tr\Big[
\epsilon^{ABCD} (\bar\Psi_A\star  X_B\star \Psi_C\star  X_D)
-\epsilon_{ABCD}( \bar\Psi^A \star X^B \star\Psi^C \star X^D )\Big],
\label{AS4a}\\
S_{4b}&=&\frac{i\kappa}{2\pi} \int d^3 x\,\tr\Big[
 \bar\Psi^A \star\Psi_A \star X_B \star X^B -  \bar\Psi_A
\star\Psi^A \star X^B \star X_B\Big],
\label{AS4b}\\
S_{4c}&=&\frac{i\kappa}{2\pi} \int d^3 x\,\tr\Big[
2 ( \bar\Psi_A\star \Psi^B\star X^A\star X_B)
- 2( \bar\Psi^A\star\Psi_B \star X_A \star X^B)\Big],
\label{AS4c}\\
S_6&=&  -\frac{1}{6}\frac{\kappa}{2\pi} \int d^3 x\,\tr(N^{IA}\star N^I_A)
\nonumber\\
&=&  \frac{1}{3}\frac{\kappa}{2\pi} \int d^3 x\,\tr \Big[
X^A\star X_A\star X^B\star X_B\star X^C\star X_C
+ X_A\star X^A\star X_B\star X^B\star X_C\star X^C
 \nonumber\\
&&\phantom{xx}
+4 X_A\star X^B\star X_C\star X^A\star X_B\star X^C
-6 X^A\star X_B\star X^B\star X_A\star X^C\star X_C
 \Big],
 \label{AS6}
\end{eqnarray}
where
\begin{eqnarray}
 N^I_A&=&\Gamma^I_{AB}\Big(X^C\star X_C\star X^B-X^B\star X_C\star X^C\Big)-2\Gamma^{IBC}X_B\star X_A\star X_C,
 \nonumber\\
 N^{IA}&=&\tilde\Gamma^{IAB}\Big(X_C\star X^C\star X_B-X_B\star X^C\star X_C\Big)-2\tilde\Gamma^{IBC}X_B\star X^A\star X_C,
\label{AS6N}
\end{eqnarray}
with $\Gamma^I_{AB}$ being $4\times 4$ matrices, the generators of the $\rm SO(6)$ group, satisfying:
\begin{eqnarray}
\Gamma^I_{AB}&=&-\Gamma^I_{BA},\,\forall I=1,...,6; \;
\Gamma^I \Gamma^J+\Gamma^J \Gamma^I=2\delta^{IJ},
\nonumber\\
\tilde\Gamma^I&=&(\Gamma^I)^\dagger\;
\Leftrightarrow\;
\tilde\Gamma^{IAB}=(\Gamma^I_{BA})^{*}=-(\Gamma^I_{AB})^{*}=
\frac{1}{2}\epsilon^{ABCD}\Gamma^I_{CD},\;
N^I_A=\big(N^{IA}\big)^\dagger.
\label{AS6N}
\end{eqnarray}
The coefficients in three possible structures for the $\Psi^2 X^2$ terms are chosen so that they give correct result required by supersymmetry. Some points are discussed and demonstrated in details in the main text and the appendix of Ref. \cite{Bandres:2008ry}.

Next we give the noncommutative gauge-fixing plus ghost terms explicitly:
\begin{eqnarray}
S_{\rm gf+ghost}=-\frac{\kappa}{2\pi} \int d^3 x \Big[\frac{1}{2\xi}\partial_\mu A^\mu\star\partial_\nu A^\nu-\bar\Lambda\star\partial_\mu D^\mu\Lambda-\frac{1}{2\xi}\partial_\mu \hat A^\mu\star\partial_\nu \hat A^\nu+\bar{\hat\Lambda}\star\partial_\mu D^\mu\hat\Lambda\Big],
\nonumber\\
\label{GfGh}
\end{eqnarray}
where covariant derivative is defined as in (\ref{CovD}): $D^\mu\Lambda=\partial\Lambda+i[A^\mu\stackrel{\star}{,} \Lambda]$.

Note that the additional interaction terms of the schematic forms $X^2\Psi^2$ and
$X^6$ are not required to deduce the equations of motion of the gauge fields, which are
\begin{equation}
J^\mu = \frac{1}{2}\epsilon^{\mu\nu\lambda} F_{\nu\lambda} \quad {\rm and} \quad \hat J^\mu =
-\frac{1}{2}\epsilon^{\mu\nu\lambda} \hat F_{\nu\lambda},
\label{EqM}
\end{equation}
where
\begin{equation}
J^\mu = iX_A D^\mu X^A -i D^\mu X_A X^A - \bar\Psi^A \gamma^\mu \Psi_A,
\label{EqM1}
\end{equation}
and
\begin{equation}
\hat J^\mu = iX^A D^\mu X_A -i D^\mu X^A X_A - \bar\Psi_A \gamma^\mu \Psi^A.
\label{EqM2}
\end{equation}
In the special case of the $\rm U(1)_{\kappa} \times U(1)_{-\kappa}$ theory one has $J^\mu = -\hat J^\mu$, and hence the equations of motion imply $F_{\mu\nu} =
\hat F_{\mu\nu}$.

\subsection{Noncommutative supersymmetry transformations}

Next, using notations of previous subsection, we give the supersymmetric transformation for the U(1) fields: $A_\mu$ and $\hat A_\mu$ gauge fields, scalar fields $X^A$, complex fermion fields $\Psi^A$, and their adjoints as well, respectively:
\begin{eqnarray}
(\delta A_\mu)^{a}{}_b&=&\Big(\Gamma^I_{AB}\bar\epsilon^I \gamma_\mu
\Psi^A\star X^B -\tilde\Gamma^{IAB}X_B\star\bar\Psi_A\gamma_\mu
\epsilon^I\Big)^{a}{}_b\;,
\nonumber\\
(\delta \hat A_\mu)^{\dot{a}}{}_{\dot{b}}&=&\Big(\Gamma^I_{AB}X^B\star\bar\epsilon^I \gamma_\mu\Psi^A  -\tilde\Gamma^{IAB}\bar\Psi_A\gamma_\mu \epsilon^I\star X_B\Big)^{\dot{a}}{}_{\dot{b}}\;,
\nonumber\\
(\delta X_A)^{a}{}_{\dot{a}}&=&\Big(i\Gamma^I_{AB}\bar\epsilon^I\star\Psi^B \Big)^{a}{}_{\dot{a}}\;,
\nonumber\\
(\delta X^A)^{\dot{a}}{}_a&=&\Big(-i \tilde\Gamma^{IAB}\bar\Psi_B\star\epsilon^I\Big)^{\dot{a}}{}_a\;,
\nonumber\\
(\delta \Psi^A)^{a}{}_{\dot{a}}&=&\Big(-\tilde\Gamma^{IAB} \gamma^\mu\epsilon^I  \star D_{\mu}X_B  + N^{IA} \star\epsilon^I \Big)^{a}{}_{\dot{a}}\;,
\nonumber\\
(\delta \Psi_A)^{\dot{a}}{}_a&=&\Big(\Gamma^I_{AB}  \gamma^{\mu}\epsilon^I
\star{D_\mu} X^B + N^I_A\star\epsilon^I \Big)^{\dot{a}}{}_a\;,
\nonumber\\
(\delta \bar\Psi_A)^{\dot{a}}{}_a&=&(\delta\Psi^T_A\gamma^0)^{\dot{a}}{}_a
=\Big(-\Gamma^I_{AB} \bar\epsilon^I \star\slashed{D} X^B + N^I_A\star\bar\epsilon^I \Big)^{\dot{a}}{}_a\;,
\label{SUSYtransf}
\end{eqnarray}
with $\bar\epsilon^I=\epsilon^I \gamma^0=(\epsilon^I)^T\gamma^0$,
and $(N^I_A)^T=N^I_A$. Detailed verification of the invariance of the NCABJM action under these transformations is presented in the appendix A.

\section{Feynman rules of the $\rm U(1)_{\kappa}\times U(1)_{-\kappa}$ NCABJM quantum field theory}

Our next task is to derive the Feynman rules needed for checking the properties of the one loop quantum corrections. In this paper we follow the usual BRST quantization, with relevant presetting  given in previous sections. We shall use a Landau gauge which amounts to the following setting of the gauge parameter: $\xi=0$, after having worked out free gauge propagators.

\begin{figure}[t]
\begin{center}
\includegraphics[width=12cm]{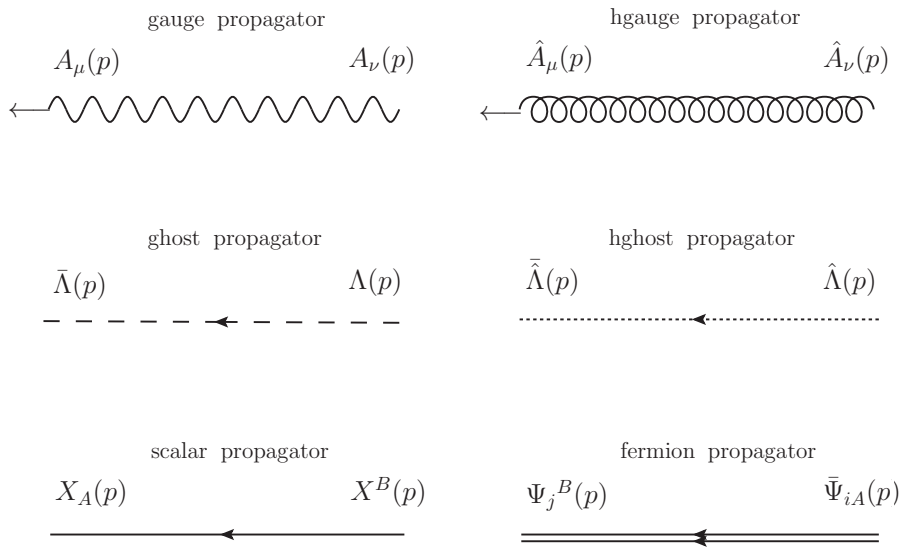}
\end{center}
\caption{Notations and the propagators of the relevant fields.}
\label{fig:notationprop1}
\end{figure}

Diagramatic notations of the relevant fields in our theory in accord with figure \ref{fig:notationprop1},
like free gauge field $A^\mu(\xi=0)$, hgauge field $\hat A^\mu(\xi=0)$, ghost $\Lambda$ and hghost $\hat\Lambda$, scalar $X^A$, and finally fermion $\psi^A_i$ field, together with their propagators in momentum space are given next, respectively:
\begin{eqnarray}
A_\nu\to A_\mu:\;\;&\Longrightarrow& \frac{2\pi}{\kappa}\Big(\frac{-\epsilon_{\mu\nu\rho}p^\rho}{p^2}\Big),\;\;\,\,
\phantom{xxx}
\hat A_\nu\to\hat A_\mu:\;\;\Longrightarrow \frac{2\pi}{\kappa}\Big(\frac{\epsilon_{\mu\nu\rho}p^\rho}{p^2}\Big),
\label{prophA}\\
\Lambda\to\bar\Lambda:\;\;&\Longrightarrow& \frac{2\pi}{\kappa}\Big(\frac{-i}{p^2}\Big),\;\;\;\,
\phantom{xxxxxxxxx}
\hat\Lambda\to\bar{\hat\Lambda}:\;\;\Longrightarrow \frac{2\pi}{\kappa}\Big(\frac{-i}{p^2}\Big),
\label{prophL}\\
X^B\to X_A:\;\;&\Longrightarrow& \frac{2\pi}{\kappa}\Big(\frac{-i}{p^2}\Big)\delta_A{}^B,\;\;\,\,
\phantom{xx.}
\bar\Psi_{Ai}\to\Psi_j{}^B:\;\;\Longrightarrow \frac{2\pi}{\kappa}\Big(\frac{-i{\slashed p}_{ij}}{p^2}\Big)\delta_A{}^B.
\label{prophS,F}
\end{eqnarray}
The interaction vertices are derived following the conventional procedure. Results are listed in the appendix D.

\section{Power counting and the limit $\theta^{\mu\nu}\rightarrow 0$}

With the relevant Feynman rules derived, we are now ready for the consistency tests of the perturbative NCABJM field theory at loop level. Before starting the computations we would like to analyze some general properties. Let's focus on an arbitrary 1PI Feynman diagram  obtained from the action (\ref{Action}) in the case of Landau gauge. Assume that the Feynman diagram in question has $E_G$ external gauge fields, $E_F$ external fermions, $E_X$ external scalars and no external ghosts. Then, it is not difficult to show that degree
of the UV divergence ${\cal D }$ for such diagram reads
\begin{equation}
{\cal D}=3-E_G-E_F-\frac{1}{2}\,E_X.
\label{5.1}
\end{equation}
Hence, all one-loop diagrams with $E_G+E_F>3$ are UV finite by power counting. Each of these diagrams is also IR finite by power counting for non-exceptional momenta, so  that one can apply Lebesgue's dominated convergence theorem and compute the limit $\theta^{\mu\nu}\to 0$ of each diagram by setting $\theta^{\mu\nu}= 0$ before the loop momentum integration. It is thus plain that all one-loop 1PI Green functions of the noncommutative ABJM quantum field theory in the Landau gauge  with   $E_G+E_F>3$ transform into the corresponding Green functions of the ordinary ABJM quantum field theory in the limit $\theta^{\mu\nu}\to 0$. The same conclusion is reached for $E_G=0=E_F$ and $E_X>6$, $E_G+E_F=1$ and $E_X=6$, $E_G+E_F=2$ and $E_X\geq 4$, and finally for $E_G+E_F=3$ and $E_X\geq 2$, respectively. However for the following combinations of triplet of number of fields: $(E_G,E_F,E_X)= (0,0,4), (1,0,4), (0,0,6), (1,0,2), (1,2,0), (0,0,2), (0,2,0)$, the power counting formula (\ref{5.1}) shows that ${\cal D}\geq 0$, i.e. it always shows the presence of UV divergence, respectively. So, the remaining 1PI Green functions fail to be UV finite by power counting and thus its limit $\theta^{\mu\nu}\to 0$ cannot be computed as we have just done. In the sections that follow, we shall work out  the limit $\theta^{\mu\nu}\to 0$ of the one-loop 1PI functions with fewer than four fields.

Let us point out that the number of scalar fields in each interaction term in the action (\ref{Action}) is even. Hence, straightforward application of Wick's theorem leads to the conclusion that any correlation function involving an odd number of scalar fields vanishes and that, if number of $X_A$ and $X^A$ fields in the correlation function is not equal it also vanishes.

\section{Gauge field $\big<A^\mu A^\nu\big>$ and hgauge field $\big<\hat A^\mu \hat A^\nu\big>$ two-point functions}

We would like to remind the reader that not all the integrals that we shall deal with in the sequel are UV finite by power-counting; so to define them and manipulate them properly, we shall use Dimensional Regularization --this is why they are defined in $D$ dimensions. Only after we have made sure that the UV divergences cancel out upon adding up contributions, we shall take the limit $D\rightarrow 3$.

Generally speaking the total $\big<A^\mu A^\nu\big>$  one-loop 1PI two-point function $\Pi^{\mu \nu}_{AA}(p)$ is the sum of the following contributions
\begin{equation}
\Pi^{\mu \nu}_{AA}(p)=(P^{\mu\nu}_{\rm bub}+P^{\mu\nu}_{\rm tad})+(F^{\mu\nu}_{\rm bub}+F^{\mu\nu}_{\rm tad})+(S^{\mu\nu}_{\rm bub}+S^{\mu\nu}_{\rm tad})
+(G^{\mu\nu}_{\rm bub}+G^{\mu\nu}_{\rm tad}),
\label{Pi}
\end{equation}
where $P^{\mu\nu}$, $F^{\mu\nu}$, $S^{\mu\nu}$, $G^{\mu\nu}$ denotes gauge field, fermion, scalar and ghost running in the bubble and/or tadpole loop, respectively. Number of contributions from (\ref{Pi}) vanish due to the absence of relevant terms in the action, i.e.
\begin{equation}
P^{\mu\nu}_{\rm tad}=F^{\mu\nu}_{\rm tad}=G^{\mu\nu}_{\rm tad}=0.
\label{Pi0}
\end{equation}
The remaining $P^{\mu\nu}_{\rm bub}$, $G^{\mu\nu}_{\rm bub}$, $S^{\mu\nu}_{\rm bub}$, $F^{\mu\nu}_{\rm bub}$, and
$S^{\mu\nu}_{\rm tad}$ we comput next.

\subsection{Gauge field bubble and tadpole diagrams}

\begin{figure}[t]
\begin{center}
\includegraphics[width=6cm]{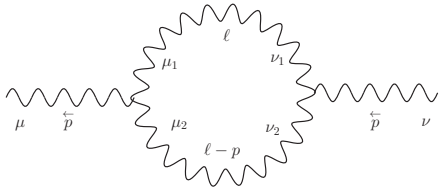}
\end{center}
\caption{Gauge field bubble-loop contribution to the gauge field 2-point function $P_{\rm bub}^{\mu\nu}$.}
\label{fig:Figl2}
\end{figure}
\begin{figure}[t]
\begin{center}
\includegraphics[width=6cm]{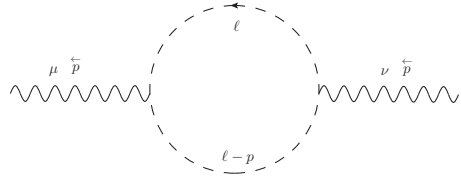}
\end{center}
\caption{Gauge field bubble, ghost-loop contribution to the 2-point function $G_{\rm bub}^{\mu\nu}$.}
\label{fig:Figl3}
\end{figure}
\begin{figure}[t]
\begin{center}
\includegraphics[width=6cm]{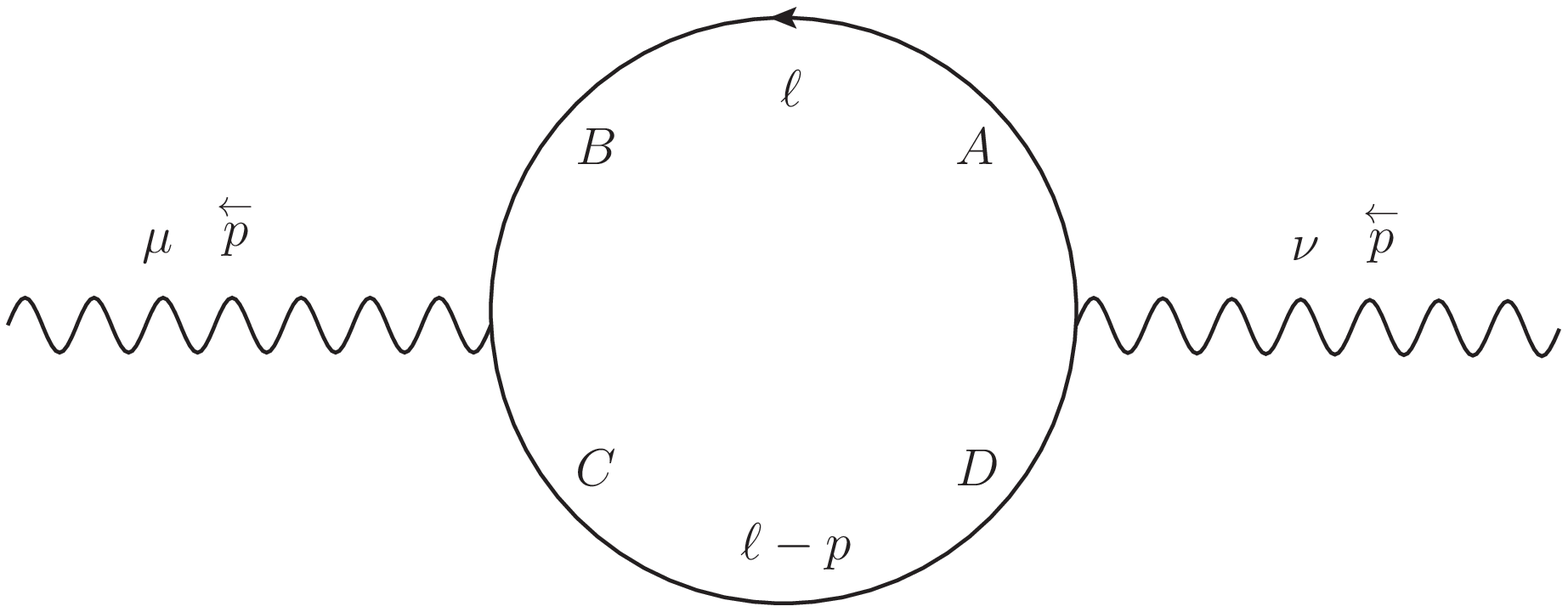}
\end{center}
\caption{Gauge field bubble, scalar-loop contribution to the 2-point function $S_{\rm bub}^{\mu\nu}$.}
\label{fig:Figl4}
\end{figure}
\begin{figure}[t]
\begin{center}
\includegraphics[width=6cm]{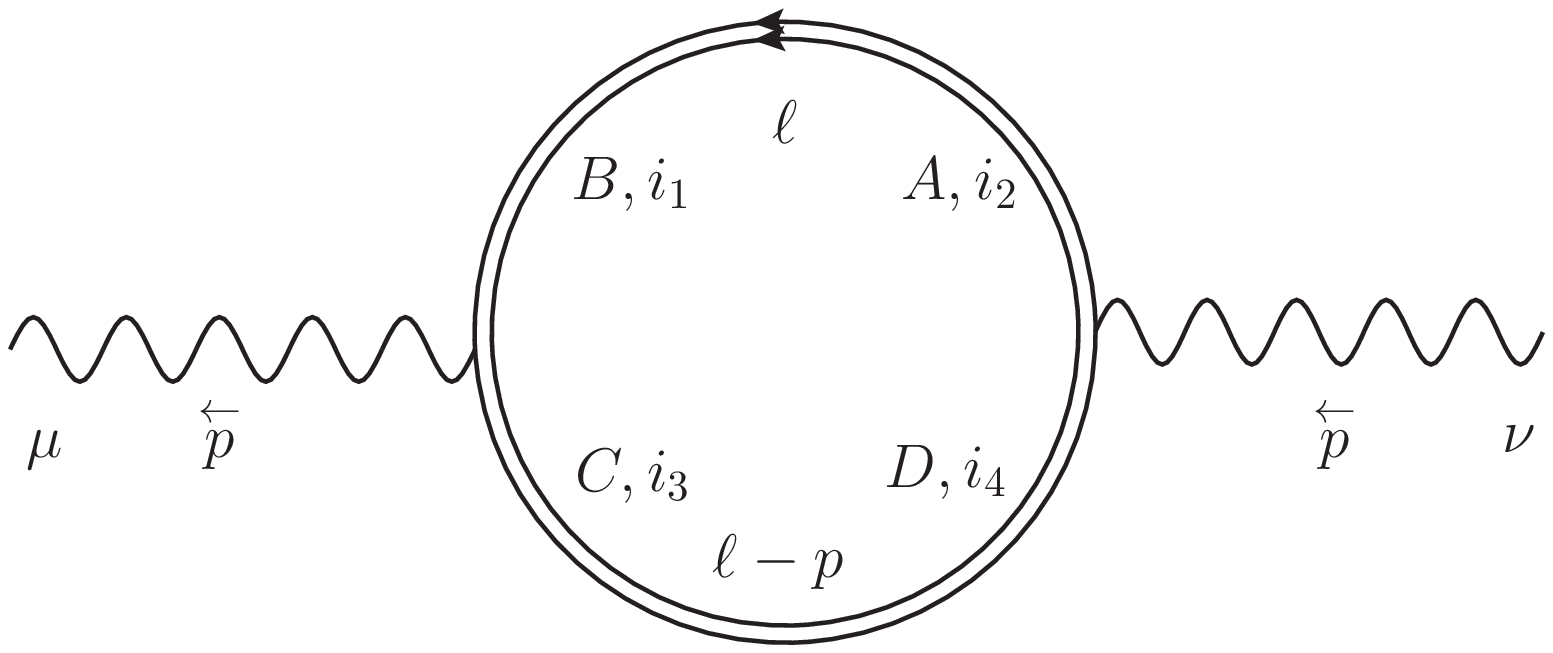}
\end{center}
\caption{Gauge field bubble, fermion-loop contribution to the 2-point function $F_{\rm bub}^{\mu\nu}$.}
\label{fig:Figl5}
\end{figure}
Using Feynman rules from the appendix D, in the appendix B we have found that contributions from the gauge field and ghost loops in the gauge field bubble diagrams, figures \ref{fig:Figl2} and \ref{fig:Figl3} respectively, are equal up to the sign:
\begin{equation}
P^{\mu\nu}_{\rm bub}=-G^{\mu\nu}_{\rm bub}=\int \frac{d^D \ell}{(2\pi)^D}
\Big(2\sin\frac{\ell\theta p}{2}\Big)^2\;\frac{\ell^\mu(\ell-p)^\nu}{\ell^2(\ell-p)^2},
\label{P+Gbub}
\end{equation}
with definition $\ell\theta p=\ell_\mu\theta^{\mu\nu}p_\nu$.

Since the phase factors cancel, contributions from scalar and fermion loops in the gauge field bubble diagrams of figures \ref{fig:Figl4} and \ref{fig:Figl5}, are:
\begin{eqnarray}
S^{\mu\nu}_{\rm bub}&=&\sum\limits_A\int \frac{d^D \ell}{(2\pi)^D}
\;\frac{4\ell^\mu\ell^\nu-2(\ell^\mu p^\nu+p^\mu\ell^\nu)+p^\mu p^\nu}{\ell^2(\ell-p)^2},
\nonumber\\
F^{\mu\nu}_{\rm bub}&=&-\sum\limits_A\int \frac{d^D \ell}{(2\pi)^D}
\;\frac{4\ell^\mu\ell^\nu-2(\ell^\mu p^\nu+p^\mu\ell^\nu)+p^2\eta^{\mu\nu}}{\ell^2(\ell-p)^2},
\nonumber\\
S^{\mu\nu}_{\rm bub}&+&F^{\mu\nu}_{\rm bub}=\sum\limits_A\,\Big(p^\mu p^\nu-p^2\eta^{\mu\nu}\Big)\int \frac{d^D \ell}{(2\pi)^D}\,\frac{1}{\ell^2(\ell-p)^2}.
\label{S+Fbub}
\end{eqnarray}

The contribution from tadpole diagram in figure \ref{fig:Figl6} vanishes:
\begin{figure}[t]
\begin{center}
\includegraphics[width=6cm]{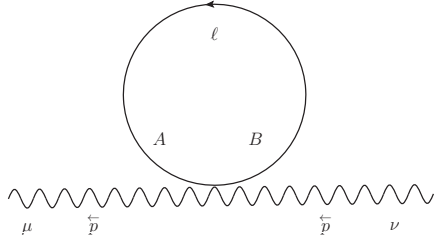}
\end{center}
\caption{Gauge field tadpole, scalar-loop contribution to the 2-point function $S^{\mu\nu}_{\rm tad}$.}
\label{fig:Figl6}
\end{figure}
\begin{equation}
S^{\mu\nu}_{\rm tad}=2\eta^{\mu\nu}\sum\limits_A\int \frac{d^D \ell}{(2\pi)^D}
\;\frac{1}{\ell^2}=0,
\label{S+Fbub1}
\end{equation}
so for $A^\mu A^\nu$ terms in the effective action we finally have the following gauge field polarization tensor:
\begin{equation}
\Pi^{\mu \nu}_{AA}(p)=S^{\mu\nu}_{\rm bub}+F^{\mu\nu}_{\rm bub}
=i\sum\limits_A\frac{1}{8}\frac{1}{\sqrt{p^2}}\Big(p^\mu p^\nu-p^2\eta^{\mu\nu}\Big).
\label{Pitot}
\end{equation}

By inspecting again Feynman rules in the appendix D it is plain that the 1PI 2-point function, $\widehat\Pi^{\mu \nu}_{\hat A \hat A}$, for the noncommutative hgauge fields from $\hat A^\mu \hat A^\nu$ terms in the action reads
\begin{equation}
\widehat\Pi^{\mu \nu}_{\hat A \hat A}(p)
=2\sum\limits_A\frac{1}{16}\frac{i}{\sqrt{p^2}}\Big(p^\mu p^\nu-p^2\eta^{\mu\nu}\Big) \equiv\Pi^{\mu \nu}_{AA}(p),
\label{6.7}
\end{equation}
so that the $\widehat\Pi^{\mu \nu}_{\hat A \hat A}(p)$ polarization tensor in the limit $\theta^{\mu\nu}\to 0$  is trivially given by the corresponding Green function --polarization tensor-- of the ordinary/commutative ABJM quantum field theory.

\section{Mixed gauge field -- hgauge field, $\big<A^\mu \hat A^\nu\big>$,  two-point functions}

For mixed $A^\mu {\hat A}^\nu$ type of terms we have the one-loop 1PI two-point function $\hat\Pi^{\mu\nu}_{A\hat A}(p)$ as a sum of contributions from figures  \ref{fig:Figl7}, \ref{fig:Figl8}, \ref{fig:Figl9}
\begin{equation}
\hat\Pi^{\mu\nu}_{A\hat A}(p)=(\hat P^{\mu\nu}_{\rm bub}+\hat P^{\mu\nu}_{\rm tad})+(\hat F^{\mu\nu}_{\rm bub}+\hat F^{\mu\nu}_{\rm tad})+(\hat S^{\mu\nu}_{\rm bub}+\hat S^{\mu\nu}_{\rm tad}).
\label{4.1hPi}
\end{equation}
Again number of contributions from (\ref{4.1hPi}) vanish due to the absence of relevant terms in the action, i.e.
\begin{equation}
\hat P^{\mu\nu}_{\rm bub}=\hat P^{\mu\nu}_{\rm tad}=F^{\mu\nu}_{\rm tad}=0.
\label{hPi0}
\end{equation}
Remaining $\hat F^{\mu\nu}_{\rm bub}$, $\hat S^{\mu\nu}_{\rm bub}$, and
$\hat S^{\mu\nu}_{\rm tad}$ we comput next.

\subsection{Gauge field -- hgauge field bubble and tadpole: scalar and fermion loops}

After some lengthy computations we found that one-loop diagrams which mix different types of gauge fields (we will call them ``mixing terms'' in discussions below) always stay non-planar (i.e. with nontrivial noncommutative phase factors). In this and next section we evaluate two- and three-point functions of this type.

\begin{figure}
\begin{center}
\includegraphics[width=6cm]{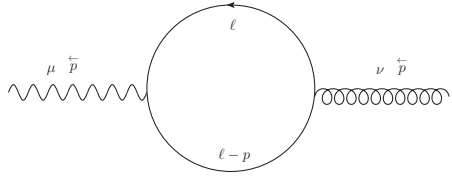}
\end{center}
\caption{Gauge field-hgauge field bubble, scalar-loop contribution to the 2-point function ${\hat S}_{\rm bub}^{\mu\nu}$.}
\label{fig:Figl7}
\end{figure}

\begin{figure}
\begin{center}
\includegraphics[width=5cm]{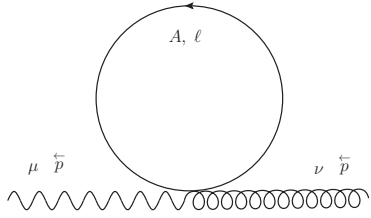}
\end{center}
\caption{Gauge field-hgauge field tadpole, scalar-loop contribution to the 2-point function ${\hat S}_{\rm tad}^{\mu\nu}$.}
\label{fig:Figl8}
\end{figure}

\begin{figure}[t]
\begin{center}
\includegraphics[width=6cm]{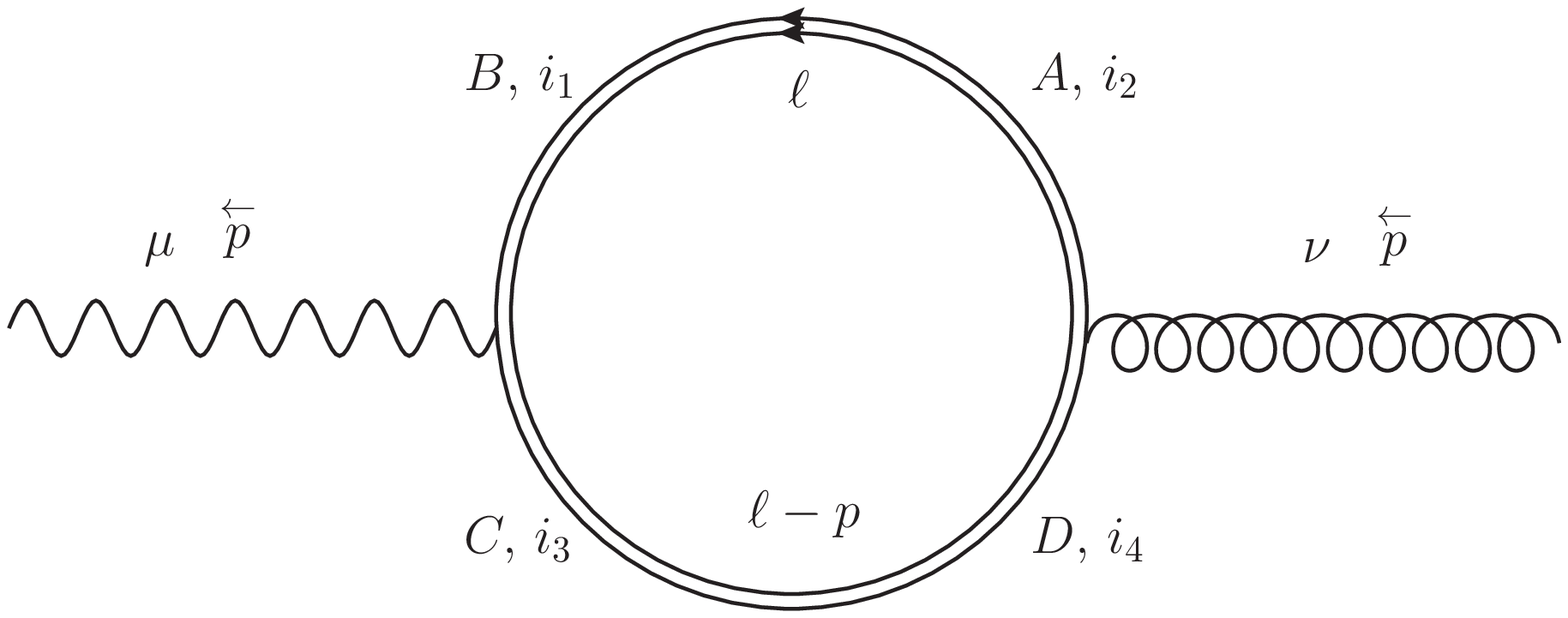}
\end{center}
\caption{Gauge field-hgauge field buble, fermion-loop contribution to the 2-point function ${\hat F}_{\rm bub}^{\mu\nu}$.}
\label{fig:Figl9}
\end{figure}

One more property of mixing terms is that they are generated by the scalar and fermion fields running in the loop only. Therefore mixed two-point function $\rm \hat\Pi_{A\hat A}$ contains three diagrams from figures \ref{fig:Figl7}, \ref{fig:Figl8}, and  \ref{fig:Figl9}:
\begin{eqnarray}
\hat S^{\mu\nu}_{\rm bub}&=&-\sum\limits_A\int \frac{d^D \ell}{(2\pi)^D}
e^{-i\ell\theta p}
\;\frac{(2\ell-p)^\mu(2\ell-p)^\nu}{\ell^2(\ell-p)^2}
\nonumber
\\
&=&-\sum\limits_A \bigg[4I_1 \eta_{\mu\nu}+\Big(4\big(I_2-I_5\big)+I\Big)p_\mu p_\nu
\nonumber\\
&&\phantom{xxxx}+2\big(2I_3-I_6\big)\big(\tilde p_\mu p_\nu+p_\mu \tilde p_\nu\big)+4I_4\tilde p_\mu \tilde p_\nu\bigg]
\nonumber
\\
&=&-\sum\limits_A \bigg[4I_1\Big(\eta_{\mu\nu}-\frac{p_\mu p_\nu}{p^2}\Big)
+\frac{i}{\pi} \frac{1}{\sqrt{\tilde p^2}}\frac{p_\mu p_\nu}{p^2}
+4I_4\tilde p_\mu \tilde p_\nu
\bigg],
\label{ShPbub}\\
\hat S^{\mu\nu}_{\rm tad}&=&2\sum\limits_A\int \frac{d^D \ell}{(2\pi)^D}
\;\frac{e^{-i\ell\theta p}}{\ell^2}\eta_{\mu\nu}
=\sum\limits_A \frac{i}{\pi}
\frac{1}{\sqrt{\tilde p^2}}\frac{\eta_{\mu\nu}}{p^2},
\label{ShPtad}\\
\hat F^{\mu\nu}_{\rm bub}&=&\sum\limits_A\int \frac{d^D \ell}{(2\pi)^D}
e^{-i\ell\theta p}
\;\frac{{\rm tr} \big(\gamma^\mu \slashed\ell\gamma^\nu(\slashed\ell-\slashed p)\big)}{\ell^2(\ell-p)^2}
\nonumber\\
&=&\sum\limits_A \bigg[\Big(4I_1 +2I p^2-\frac{i}{2\pi}\frac{1}{\sqrt{\tilde p^2}}\Big)\Big(\eta_{\mu\nu}-\frac{p_\mu p_\nu}{p^2}\Big)+4I_4\tilde p_\mu \tilde p_\nu
\bigg],
\label{FhPbub}
\end{eqnarray}
where we denote two structures $k_\mu\theta^{\mu\nu}p_\nu=k\theta p$ and $\tilde p^\mu=\theta^{\mu\nu}p_\nu$, respectively. For the definitions and details of the above integrals $I,I_1,....., I_6$, see the appendix C.

Once we sum over all contributions and perform a standard tensor reduction, the integral boils down to a single tensor structure multiplying one scalar master integral $I(p,\theta)$, which in the Minkovski signature is $I^M(p,\theta)$. So, from mixed $A^\mu \hat A^\nu$ terms we finally have the following polarization tensor: \begin{eqnarray}
\hat\Pi^{\mu\nu}_{A\hat A}=\hat S^{\mu\nu}_{\rm bub}+
\hat S^{\mu\nu}_{\rm tad} + \hat F^{\mu\nu}_{\rm bub}
&=&\sum\limits_A\big(p^2\eta^{\mu\nu}-p^\mu p^\nu\big)\int\frac{d^D\ell}{(2\pi)^D}\frac{e^{-i\ell\theta p}}{\ell^2(\ell-p)^2}
\nonumber\\
&=&\sum\limits_A \big(p^2\eta^{\mu\nu}-p^\mu p^\nu\big)I^M(p,\theta),
\label{hPitot}
\end{eqnarray}
with $I^M(p,\theta)$ for Minkowski signature being given in the appendix C by (\ref{C.15}) via (\ref{C.14}). Taking commutative limit $\theta^{\mu\nu}\to 0$ the above polarization tensor $i\hat\Pi^{\mu\nu}_{A\hat A}$ from (\ref{hPitot}) takes very simple form:
\begin{equation}
\lim_{\theta\to 0}i\hat\Pi^{\mu\nu}_{A\hat A}=\sum\limits_A \big(p^2\eta^{\mu\nu}-p^\mu p^\nu\big) \lim_{\theta\to 0}I^M(p,\theta)
=i\sum\limits_A\frac{\big(p^2\eta^{\mu\nu}-p^\mu p^\nu\big)}{8\sqrt{p^2-i0^+}},
\label{hPitot0}
\end{equation}
i.e. $I^M(p,\theta)$ clearly converges to the commutative value smoothly when
$\theta^{\mu\nu}\to 0$, which is precisely the 1-loop contribution to the $i\hat\Pi^{\mu\nu}_{A\hat A}$ in the ordinary/commutative ABJM theory.


\section{Gauge field $\big<A^{\mu_1} A^{\mu_2} A^{\mu_3}\big>$ and hgauge field $\big<\hat A^{\mu_1} \hat A^{\mu_2} \hat A^{\mu_3}\big>$, three-point functions}

\begin{figure}[t]
\begin{center}
\includegraphics[width=12cm]{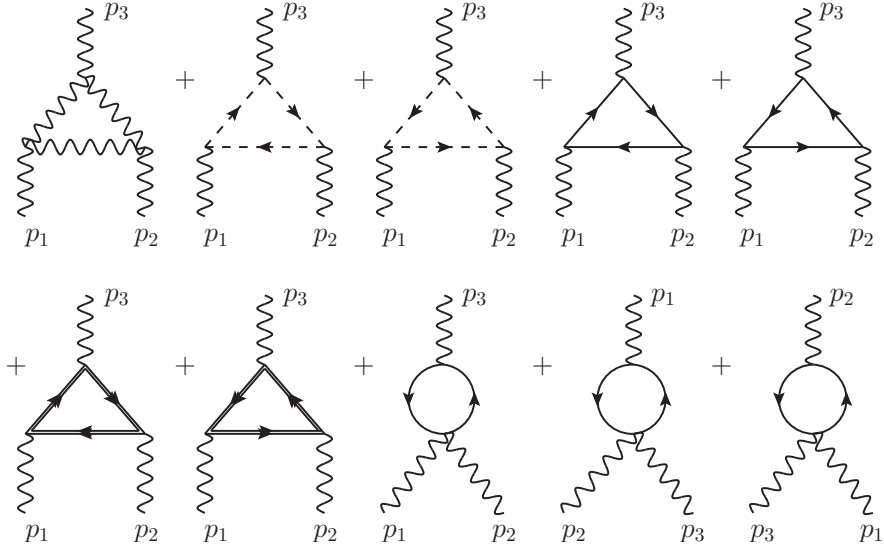}
\end{center}
\caption{One-loop contributions to the gauge field 3-point function $\big<A^{\mu_1} A^{\mu_2} A^{\mu_3}\big>$.}
\label{fig:Figl10}
\end{figure}

From Feynmanm rules in Appendix D 
we have one-loop 1PI three-point function $\Pi^{\mu_1\mu_2\mu_3}_{AAA}$ as a sum of contributions from diagrams in figure \ref{fig:Figl10},
\begin{eqnarray}
\Pi^{\mu_1\mu_2\mu_3}_{AAA}&=&
P^{\mu_1\mu_2\mu_3}_{\rm tria1}+G^{\mu_1\mu_2\mu_3}_{\rm tria1}+
G^{\mu_1\mu_2\mu_3}_{\rm tria2}+S^{\mu_1\mu_2\mu_3}_{\rm tria1}+S^{\mu_1\mu_2\mu_3}_{\rm tria2}
\nonumber\\
&+&
F^{\mu_1\mu_2\mu_3}_{\rm tria1}+F^{\mu_1\mu_2\mu_3}_{\rm tria2}+
S^{\mu_1\mu_2\mu_3}_{\rm bub1}+S^{\mu_1\mu_2\mu_3}_{\rm bub2}+ S^{\mu_1\mu_2\mu_3}_{\rm bub3},
\label{AAAPi}
\end{eqnarray}
while for $\hat A^{\mu_1} \hat A^{\mu_2} \hat A^{\mu_3}$ terms in the effective action $S$ (\ref{Action}) we have the one-loop 1PI three-point functions $\widehat\Pi^{\mu_1\mu_2\mu_3}_{\hat A\hat A\hat A}$ as a sum of contributions from the sum of diagrams in figure \ref{fig:Figl10} where all {\it wavy} gauge field lines are replaced by the {\it curly} hgauge field lines with relevant Feynman rules given in the appendix D, for every pair $(\mu_i,p_i),\, i=1,2,3$:
\begin{eqnarray}
\widehat\Pi^{\mu_1\mu_2\mu_3}_{\hat A\hat A\hat A}&=&
\widehat P^{\mu_1\mu_2\mu_3}_{\rm tria1}+\widehat G^{\mu_1\mu_2\mu_3}_{\rm tria2}+
\widehat G^{\mu_1\mu_2\mu_3}_{\rm bub1}+\widehat S^{\mu_1\mu_2\mu_3}_{\rm tria1}
+\widehat S^{\mu_1\mu_2\mu_3}_{\rm tria2}
\nonumber\\
&+&
\widehat F^{\mu_1\mu_2\mu_3}_{\rm tria1}+\widehat F^{\mu_1\mu_2\mu_3}_{\rm tria2}+
\widehat S^{\mu_1\mu_2\mu_3}_{\rm bub1}+\widehat S^{\mu_1\mu_2\mu_3}_{\rm bub2}
+ \widehat S^{\mu_1\mu_2\mu_3}_{\rm bub3}.
\label{AAAhhhPi}
\end{eqnarray}
In Eqs. (\ref{AAAPi}) and (\ref{AAAhhhPi}), $P,G,S$ and $F$ denote gauge field, ghost, scalar and fermion loops, respectivly. Other contributions vanish due to the absence of relevant terms in the action (\ref{Action}). Remaining non-vanishing terms in (\ref{AAAPi}) and (\ref{AAAhhhPi}) are presented next by looking into the one-loop corrections to the identical three gauge field vertex.

There are three relevant diagrams: the gauge field triangle (1st diagram in figure \ref{fig:Figl10}) and clockwise/counterclockwise running loop-momenta ghost triangles (2nd and 3d diagrams in figure \ref{fig:Figl10}), contributing to the $P^{\mu_1\mu_2\mu_3}_{\rm tria1}$, $G^{\mu_1\mu_2\mu_3}_{\rm tria1}$, and to the $G^{\mu_1\mu_2\mu_3}_{\rm tria2}$, respectively.\\
The gauge field triangle is as follows:
\begin{equation}
\begin{split}
P^{\mu_1\mu_2\mu_3}_{\rm tria1}=&\int\frac{d^D\ell}{(2\pi)^D} (-2i)^3\sin\frac{-\ell\theta(\ell-p_2)}{2}\sin\frac{(-\ell+p_1)\theta(\ell-p_1-p_2))}{2}
\\&
\cdot\sin\frac{(-\ell+p_1+p_2)\theta\ell}{2}
\epsilon^{\sigma_2\mu_1\sigma_1}\epsilon_{\sigma_1\rho_1\sigma_3}\epsilon^{\sigma_3\mu_2\sigma_4}\epsilon_{\sigma_4\rho_2\sigma_5}\epsilon^{\sigma_5\mu_3\sigma_6}\epsilon_{\sigma_6\rho_3\sigma_2}
\\&
\cdot\frac{\ell^{\rho_1}(\ell-p_1-p_2)^{\rho_2}(\ell-p_1)^{\rho_3}}{\ell^2(\ell-p_1)^2(\ell-p_1-p_2)^2},
\end{split}
\label{F.1}
\end{equation}
while the ghost triangles read:
\begin{equation}
\begin{split}
G^{\mu_1\mu_2\mu_3}_{\rm tria1}=&\int\frac{d^D\ell}{(2\pi)^D}(2i)^3\sin\frac{p_1\theta\ell}{2}\sin\frac{p_2\theta(\ell-p_1)}{2}\sin\frac{(-p_1-p_2)\theta(\ell-p_1-p_2)}{2}
\\&\cdot\frac{(\ell-p_1)^{\mu_1}(\ell-p_1-p_2)^{\mu_2}\ell^{\mu_3}}{\ell^2(\ell-p_1)^2(\ell-p_1-p_2)^2},
\end{split}
\label{F.2}
\end{equation}
\begin{equation}
\begin{split}
G^{\mu_1\mu_2\mu_3}_{\rm tria2}=&\int\frac{d^D\ell}{(2\pi)^D}(2i)^3\sin\frac{p_2\theta\ell}{2}\sin\frac{p_1\theta(\ell-p_2)}{2}\sin\frac{(-p_1-p_2)\theta(\ell-p_1-p_2)}{2}
\\&\cdot\frac{(\ell-p_2)^{\mu_2}(\ell-p_1-p_2)^{\mu_1}\ell^{\mu_3}}{\ell^2(\ell-p_2)^2(\ell-p_1-p_2)^2}.
\end{split}
\label{F.3}
\end{equation}
Using a simple transformation $\ell\to-\ell+p_1+p_2$ one can turn the denominator and the phase factor of the $G^{\mu_1\mu_2\mu_3}_{\rm tria2}$ to be identical to those in $G^{\mu_1\mu_2\mu_3}_{\rm tria1}$,
\begin{equation}
\begin{split}
G^{\mu_1\mu_2\mu_3}_{\rm tria1}=&\int\frac{d^D\ell}{(2\pi)^D}(2i)^3\sin\frac{p_1\theta\ell}{2}\sin\frac{p_2\theta(\ell-p_1)}{2}\sin\frac{(-p_1-p_2)\theta(\ell-p_1-p_2)}{2}
\\&\cdot\frac{(\ell-p_1)^{\mu_2}(\ell-p_1-p_2)^{\mu_3}\ell^{\mu_1}}{\ell^2(\ell-p_1)^2(\ell-p_1-p_2)^2}.
\end{split}
\label{F.4}
\end{equation}
Summing over $P^{\mu_1\mu_2\mu_3}_{\rm tria1}$, $G^{\mu_1\mu_2\mu_3}_{\rm tria1}$ and $G^{\mu_1\mu_2\mu_3}_{\rm tria2}$, and reducing the Levi-Civita symbols into metric contractions, we get
\begin{equation}
\begin{split}
P^{\mu_1\mu_2\mu_3}_{\rm tria1}+&G^{\mu_1\mu_2\mu_3}_{\rm tria1}+G^{\mu_1\mu_2\mu_3}_{\rm tria2}=-8i\int\frac{d^D\ell}{(2\pi)^D}\sin\frac{\ell\theta p_1}{2}\sin\frac{(\ell-p_1)\theta p_2}{2}\sin\frac{\ell\theta(p_1+p_2)}{2}
\\&\cdot\frac{\ell^{\mu_2}(p_1^{\mu_3}p_2^{\mu_1}-p_1^{\mu_1}p_2^{\mu_3})+\ell^{\mu_3}(p_1^{\mu_2}p_2^{\mu_1}-p_2^{\mu_2}p_1^{\mu_1})+\ell^{\mu_1}(p_1^{\mu_2}p_2^{\mu_3}-p_1^{\mu_3}p_2^{\mu_2})}{\ell^2(\ell-p_1)^2(\ell-p_1-p_2)^2}.
\end{split}
\label{F.5}
\end{equation}
If one removes the $\sin$ functions from the integrand of the previous integral, one ends up with an integral which is both UV and IR divergent by power counting. Hence, one can apply Lebesgue's dominated convergence theorem and commute the limit
$\theta\rightarrow 0$ with the integral symbol in (\ref{F.5}) to conclude that
\begin{equation}
\lim_{\theta\rightarrow 0} \Big[P^{\mu_1\mu_2\mu_3}_{\rm tria1}+G^{\mu_1\mu_2\mu_3}_{\rm tria1}+G^{\mu_1\mu_2\mu_3}_{\rm tria2}\Big]=0.
\label{8.8}
\end{equation}
This is in the full agreement with the fact that in the ordinary abelian ABJM field theory the first three Feynman diagrams from figure \ref{fig:Figl10} do not exist.

Now, by using Feynman rules one can easy show that the last seven diagrams in figure \ref{fig:Figl10} do not involve nonplanar contributions, i.e., the Moyal phases in them do not involve the loop momentum but only the external momenta. Hence the limit $\theta^{\mu\nu}\to 0$ exists trivially at $D=3$, and, if sum of those seven diagrams is UV finite for nonzero $\theta^{\mu\nu}$, it is  given by the ordinary result. One can show that this is the case. Indeed, the sum of the contributions to the 4th and 5th diagrams which are not UV finite by power counting reads
\begin{equation}
-\sum\limits_A\big(
e^{\frac{i}{2}p_1\theta p_2}+e^{-\frac{i}{2}p_1\theta p_2}\big)
\int \frac{d^3 \ell}{(2\pi)^3} \frac{8\,\ell^{\mu_1}\ell^{\mu_2}\ell^{\mu_3}}{\ell^2(\ell+p_1)^2(\ell+p_1+p_2)^2}.
\label{zerouno}
\end{equation}

It can be shown that the sum of contributions to the 6th and 7th diagrams which are not UV finite by power counting is given by
\begin{equation}
\sum\limits_A\,\big(
e^{\frac{i}{2}p_1\theta p_2}+e^{-\frac{i}{2}p_1\theta p_2}\big)
\int \frac{d^3 \ell}{(2\pi)^3} \frac{8\,\ell^{\mu_1}\ell^{\mu_2}\ell^{\mu_3}-2 \ell^2(\ell^{\mu_1}\eta^{\mu_2\mu_3}+\ell^{\mu_2}\eta^{\mu_1\mu_3}+\ell^{\mu_3}\eta^{\mu_1\mu_3})
}{\ell^2(\ell+p_1)^2(\ell+p_1+p_2)^2}.
\label{zerodos}
\end{equation}

By adding contributions of the last three diagrams in figure \ref{fig:Figl10}, which are not UV finite by power counting, one obtains
\begin{equation}
\sum\limits_A\,\big(
e^{\frac{i}{2}p_1\theta p_2}+e^{-\frac{i}{2}p_1\theta p_2}\big)
\int \frac{d^3 \ell}{(2\pi)^3} \frac{2\ell^2(\ell^{\mu_1}\eta^{\mu_2\mu_3}+\ell^{\mu_2}\eta^{\mu_1\mu_3}+\ell^{\mu_3}\eta^{\mu_1\mu_3})
}{\ell^2(\ell+p_1)^2(\ell+p_1+p_2)^2}.
\label{zerotres}
\end{equation}
Finally, the sum of equations (\ref{zerouno}), (\ref{zerodos}) and (\ref{zerotres}) is plain zero. Hence, the sum of the last seven diagram of  figure \ref{fig:Figl10} is indeed UV finite by power counting for non-zero $\theta^{\mu\nu}$, so that its $\theta^{\mu\nu}\to 0$ limit is given by the corresponding sum of diagrams of the commutative ABJM theory.

In summary, we have shown that the sum of all diagrams in figure \ref{fig:Figl10} involves only integrals which are UV finite by power counting and that the limit $\theta^{\mu\nu}\rightarrow 0$ of the sum is given by the sum of relevant diagrams in the ordinary ABJM field theory. Hence, the one-loop 1PI contribution to the
$\big<A^{\mu_1} A^{\mu_2} A^{\mu_3}\big>$ is UV finite and by taking the limit $\theta^{\mu\nu}\rightarrow 0$ of it one obtains the corresponding Green function of the ordinary ABJM quantum field theory. From Feynman rules in the appendix D it is clear that the same holds for the $\big<\hat A^{\mu_1} \hat A^{\mu_2} \hat A^{\mu_3}\big>$ three-point function.

\section{Mixed gauge field -- hgauge field, $\big<A^{\mu_1} A^{\mu_2} \hat A^{\mu_3}\big>$, $\big<\hat A^{\mu_1} \hat A^{\mu_2} A^{\mu_3}\big>$, three-point functions}

For mixed $A^{\mu_1} A^{\mu_2} \hat A^{\mu_3}$ type of terms we have the one-loop three-point function $\hat\Pi^{\mu_1\mu_2\mu_3}_{AA\hat A}$ as a sum of seven contributions, two from clockwise and counterclockwise running scalars, three running scalars in bubbles and two fermion clockwise and counterclockwise triangles as shown in figures \ref{fig:Figl11}, \ref{fig:Figl12}, \ref{fig:Figl13}, \ref{fig:Figl14}, \ref{fig:Figl15}, \ref{fig:Figl16} and \ref{fig:Figl17}. We denote them as follows, respectively:
\begin{equation}
\hat\Pi^{\mu_1\mu_2\mu_3}_{AA\hat A}=
\hat S^{\mu_1\mu_2\mu_3}_{\rm tria1}+\hat S^{\mu_1\mu_2\mu_3}_{\rm tria2}+
\hat S^{\mu_1\mu_2\mu_3}_{\rm bub1}+\hat S^{\mu_1\mu_2\mu_3}_{\rm bub2}+\hat S^{\mu_1\mu_2\mu_3}_{\rm bub3}+
\hat F^{\mu_1\mu_2\mu_3}_{\rm tria1}+\hat F^{\mu_1\mu_2\mu_3}_{\rm tria2}.
\label{AAAhPi}
\end{equation}
Other contributions vanish due to the absence of relevant terms in the action. 

For $\hat A^{\mu_1} \hat A^{\mu_2} A^{\mu_3}$ type of terms we have the one-loop 1PI three-point function $\tilde\Pi^{\mu_1\mu_2\mu_3}_{\hat A\hat A A}$ as a sum of contributions from the same figures \ref{fig:Figl11}, \ref{fig:Figl12}, \ref{fig:Figl13}, \ref{fig:Figl14}, \ref{fig:Figl15}, \ref{fig:Figl16} and \ref{fig:Figl17}, where the {\it wavy} gauge field lines are replaced with {\it curly} gauge field lines and vice-versa ({\it wavy $\leftrightarrow$ curly}).
\begin{equation}
\tilde\Pi^{\mu_1\mu_2\mu_3}_{\hat A\hat A A}=
{\tilde S}^{\mu_1\mu_2\mu_3}_{\rm tria1}+{\tilde S}^{\mu_1\mu_2\mu_3}_{\rm tria2}+
{\tilde S}^{\mu_1\mu_2\mu_3}_{\rm bub1}+{\tilde S}^{\mu_1\mu_2\mu_3}_{\rm bub2}+{\tilde S}^{\mu_1\mu_2\mu_3}_{\rm bub3}+ {\tilde F}^{\mu_1\mu_2\mu_3}_{\rm tria1}+{\tilde F}^{\mu_1\mu_2\mu_3}_{\rm tria2}.
\label{AAAhhPi}
\end{equation}
Other contributions vanish due to the absence of relevant terms in the action. Remaining terms in (\ref{AAAhPi}) and (\ref{AAAhhPi}) we compute next by using Feynman rules from the appendix D.

\subsection{Loop integrals contributing to the $\big<A^{\mu_1} A^{\mu_2} \hat A^{\mu_3}\big>$ 3-point function}

\begin{figure}[t]
\begin{center}
\includegraphics[width=5cm]{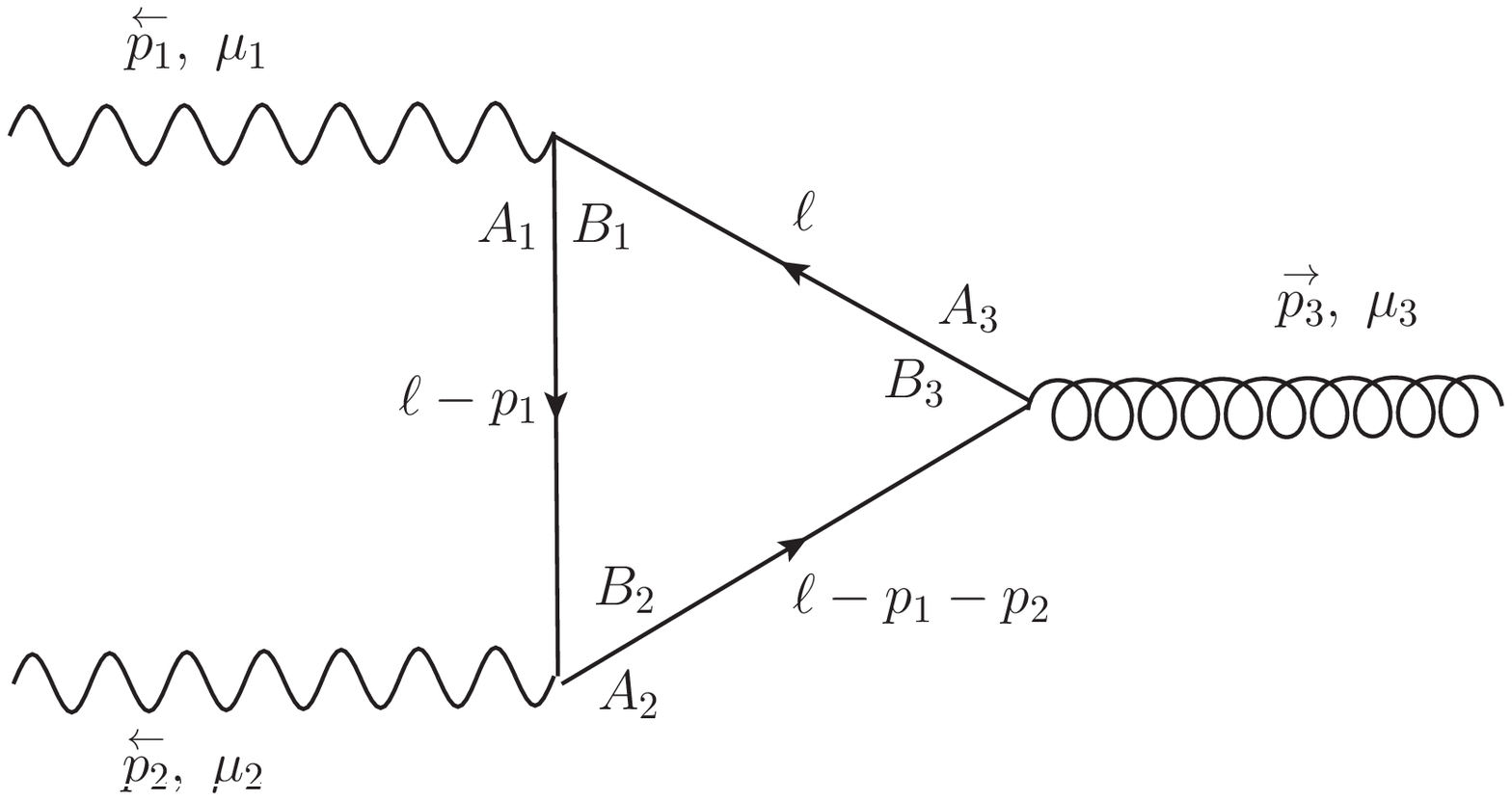}
\end{center}
\caption{Scalar triangle-loop contribution1 to the 3-point function $\hat S^{\mu_1\mu_2\mu_3}_{\rm tria1}$.}
\label{fig:Figl11}
\end{figure}

\begin{figure}[t]
\begin{center}
\includegraphics[width=5cm]{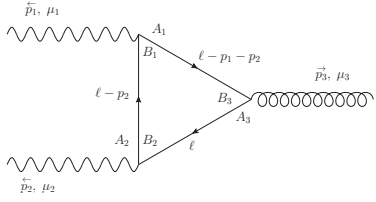}
\end{center}
\caption{Scalar triangle-loop contribution2 to the 3-point function $\hat S^{\mu_1\mu_2\mu_3}_{\rm tria2}$.}
\label{fig:Figl12}
\end{figure}

Computation of Fynman diagrams from figures \ref{fig:Figl11} and \ref{fig:Figl12}, gives, respectively:
\begin{eqnarray}
\hat S^{\mu_1\mu_2\mu_3}_{\rm tria1}&=&\sum\limits_A
e^{\frac{i}{2}p_1\theta p_2}
\int \frac{d^D \ell}{(2\pi)^D}
e^{-i\ell\theta (p_1+p_2)}
\nonumber\\
&&\phantom{XXXXXXXXXX}\cdot\frac{(2\ell-p_1)^{\mu_1}(2\ell-2p_1-p_2)^{\mu_3}(2\ell-p_1-p_2)^{\mu_2}}{\ell^2(\ell-p_1)^2(\ell-p_1-p_2)^2},
\label{Striph(-)}\\
\hat S^{\mu_1\mu_2\mu_3}_{\rm tria2}&=&\sum\limits_A
e^{-\frac{i}{2}p_1\theta p_2}
\int \frac{d^D \ell}{(2\pi)^D}
e^{-i\ell\theta(p_1+p_2)}
\nonumber\\
&&\phantom{XXXXXXXXXX}\cdot \frac{(2\ell-p_2)^{\mu_2}(2\ell-p_1-2p_2)^{\mu_3}(2\ell-p_1-p_2)^{\mu_1}}{\ell^2(\ell-p_2)^2(\ell-p_1-p_2)^2}.
\label{Striph(+)}
\end{eqnarray}
Inspecting eqs. (\ref{Striph(-)}) and (\ref{Striph(+)}) one finds out that diagrams in figures \ref{fig:Figl11} and \ref{fig:Figl12} transfer one to each other by simple replacement:\begin{equation}
\hat S^{\mu_1\mu_2\mu_3}_{\rm tria1}
\bigg|_{\substack{p_1\leftrightarrow p_2\\ \mu_1\leftrightarrow \mu_2}}=
\hat S^{\mu_1\mu_2\mu_3}_{\rm tria2}.
\label{Stri12}
\end{equation}


\begin{figure}[t]
\begin{center}
\includegraphics[width=5cm]{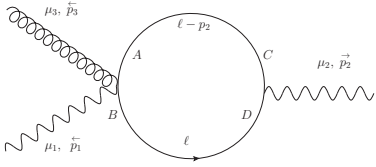}
\end{center}
\caption{Scalar bubble-loop contribution1 to the 3-point function $\hat S^{\mu_1\mu_2\mu_3}_{\rm bub1}$.}
\label{fig:Figl13}
\end{figure}

\begin{figure}[t]
\begin{center}
\includegraphics[width=5cm]{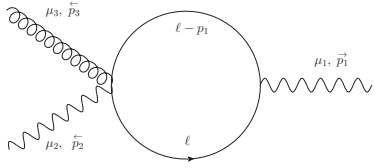}
\end{center}
\caption{Scalar bubble-loop contribution2 to the 3-point function $\hat S^{\mu_1\mu_2\mu_3}_{\rm bub2}$.}
\label{fig:Figl14}
\end{figure}

\begin{figure}[t]
\begin{center}
\includegraphics[width=5cm]{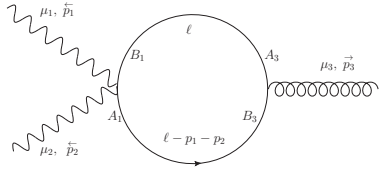}
\end{center}
\caption{Scalar bubble-loop contribution3 to the 3-point function $\hat S^{\mu_1\mu_2\mu_3}_{\rm bub3}$.}
\label{fig:Figl15}
\end{figure}

\begin{figure}[t]
\begin{center}
\includegraphics[width=5cm]{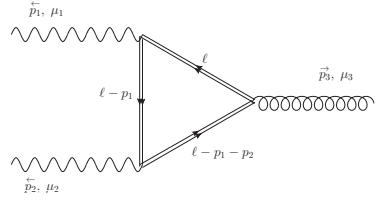}
\end{center}
\caption{Fermion triangle-loop contribution1 to the 3-point function $\hat F^{\mu_1\mu_2\mu_3}_{\rm tria1}$.}
\label{fig:Figl16}
\end{figure}

\begin{figure}[t]
\begin{center}
\includegraphics[width=5cm]{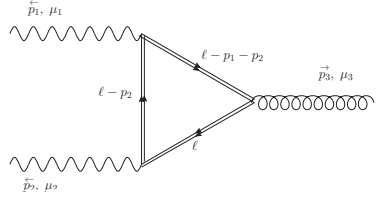}
\end{center}
\caption{Fermion triangle-loop contribution2 to the 3-point function $\hat F^{\mu_1\mu_2\mu_3}_{\rm tria2}$.}
\label{fig:Figl17}
\end{figure}

From Fynman diagrams in figures \ref{fig:Figl13} and \ref{fig:Figl14} we have
\begin{eqnarray}
\hat S^{\mu_1\mu_2\mu_3}_{\rm bub1}&=&-\sum\limits_A
e^{\frac{i}{2}p_1\theta p_2}
\int \frac{d^D \ell}{(2\pi)^D}
e^{-i\ell\theta(p_1+p_2)}
\frac{\eta^{\mu_2\mu_3}(2\ell-p_1)^{\mu_1}}{\ell^2(\ell-p_1)^2},
\nonumber\\
\label{Stribub2}
\end{eqnarray}
\begin{eqnarray}
\hat S^{\mu_1\mu_2\mu_3}_{\rm bub2}&=&-\sum\limits_A
e^{-\frac{i}{2}p_1\theta p_2}
\int \frac{d^D \ell}{(2\pi)^D}
e^{-i\ell\theta(p_1+p_2)}
\frac{\eta^{\mu_1\mu_3}(2\ell-p_2)^{\mu_2}}{\ell^2(\ell-p_2)^2},
\nonumber\\
\label{Stribub2}
\end{eqnarray}
\noindent
while in diagram from figure \ref{fig:Figl15} two phase terms combine into the 
$\cos$ function of external momenta:
\begin{eqnarray}
\hat S^{\mu_1\mu_2\mu_3}_{\rm bub3}
&=&\hat S^{\mu_1\mu_2\mu_3}_{\rm bub3+}+\hat S^{\mu_1\mu_2\mu_3}_{\rm bub3-}
\nonumber\\
&=&
-2\cos{\frac{p_1\theta p_2}{2}}\sum\limits_A
\int \frac{d^D \ell}{(2\pi)^D}
e^{-i\ell\theta(p_1+p_2)}
\frac{\eta^{\mu_1\mu_2}(2\ell-p_1-p_2)^{\mu_3}}{\ell^2(\ell-p_1-p_2)^2}.
\label{Stribub3}
\end{eqnarray}

Finally computation of Fynman diagrams from figures \ref{fig:Figl16} and \ref{fig:Figl17}, for $D=3$ gives:
\begin{eqnarray}
\hat F^{\mu_1\mu_2\mu_3}_{\rm tria1}&=&-\sum\limits_A
e^{\frac{i}{2}p_1\theta p_2}\int\frac{d^D\ell}{(2\pi)^D}e^{-i\ell\theta (p_1+p_2)}\frac{\tr\gamma^{\mu_1}\slashed{\ell}\gamma^{\mu_3}(\slashed{\ell}-\slashed{p}_1-\slashed{p}_2)\gamma^{\mu_2}(\slashed\ell-\slashed{p}_1)}{\ell^2(\ell-p_1)^2(\ell-p_1-p_2)^2},
\nonumber\\
\label{Ftriph1}\\
\hat F^{\mu_2\mu_2\mu_3}_{\rm tria2}&=&-\sum\limits_A
e^{-\frac{i}{2}p_1\theta p_2}\int\frac{d^D\ell}{(2\pi)^D}e^{-i\ell\theta (p_1+p_2)}\frac{\tr\gamma^{\mu_2}\slashed{\ell}\gamma^{\mu_3}(\slashed{\ell}-\slashed{p}_1-\slashed{p}_2)\gamma^{\mu_1}(\slashed\ell-\slashed{p}_2)}{\ell^2(\ell-p_2)^2(\ell-p_1-p_2)^2}.
\nonumber\\
\label{Ftriph2}
\end{eqnarray}

\subsection{Computations of the $\big<A^{\mu_1} A^{\mu_2} \hat A^{\mu_3}\big>$ 3-point functions}

Performing some computations of diagrams in figures \ref{fig:Figl11},  \ref{fig:Figl12}, \ref{fig:Figl16} and \ref{fig:Figl17} we find that the opposite loop momenta running generates opposite overall phase factors $e^{\pm\frac{i}{2}p_1\theta p_2 }$. We then use such phases to decompose the rest of the tensor $\hat\Pi^{\mu_1\mu_2\mu_3}_{AA\hat A}$ into two groups, the $\hat\Pi_{AA\hat A+}^{\mu_1\mu_2\mu_3}$ and the $\hat\Pi_{AA\hat A-}^{\mu_1\mu_2\mu_3}$ tensors, respectively. There we have three terms from scalar bubble diagrams, two of them carry fixed running phase while the last one contains two terms with opposite phases, they are marked as well as the two scalar triangles and the two fermion triangles.

Next step is to sum over contributions to each phase, for the clockwise running part:
\begin{equation}
\hat\Pi_{AA\hat A+}^{\mu_1\mu_2\mu_3}=\hat S^{\mu_1\mu_2\mu_3}_{\rm tria1}+\hat F^{\mu_1\mu_2\mu_3}_{\rm tria1}+\hat S^{\mu_1\mu_2\mu_3}_{\rm bub1}+\hat S^{\mu_1\mu_2\mu_3}_{\rm bub3+},
\label{9.11}
\end{equation}
and, for the counterclockwise running part:
\begin{equation}
\hat\Pi_{AA\hat A-}^{\mu_1\mu_2\mu_3}=\hat S^{\mu_1\mu_2\mu_3}_{\rm tria2}+\hat F^{\mu_1\mu_2\mu_3}_{\rm tria2}+\hat S^{\mu_1\mu_2\mu_3}_{\rm bub2}+\hat S^{\mu_1\mu_2\mu_3}_{\rm bub3-}.
\label{9.12}
\end{equation}
After summing over all terms with loop momenta carrying more than one external index, i.e. $\ell^{\mu_1}\ell^{\mu_2}\ell^{\mu_3}$ and $\ell^{\mu_i}\ell^{\mu_j}, i,j=1,2,3$ terms, cancel. Now we use the standard relation $2\ell\cdot p=(\ell+p)^2-p^2-\ell^2$ to turn the higher power in $\ell$ terms in the triangle integral into the bubble type of integrals. We also observe that
\begin{equation}
\int\frac{d^D\ell}{(2\pi)^D}\frac{e^{\frac{i}{2}p_1\theta p_2}e^{i\ell\theta (p_1+p_2)}}{(\ell-p_1)^2(\ell-p_1-p_2)^2}=\int\frac{d^D\ell}{(2\pi)^D}\frac{e^{-\frac{i}{2}p_1\theta p_2}e^{i\ell\theta (p_1+p_2)}}{\ell^2(\ell-p_2)^2},
\label{9.13}
\end{equation}
therefore such terms after the transformation have to be moved from one group to the other, and than, as indicated above, the tensor $\hat\Pi^{\mu_1\mu_2\mu_3}_{AA\hat A}$ boils down to
\begin{equation}
\hat\Pi^{\mu_1\mu_2\mu_3}_{AA\hat A}=\hat\Pi_{AA\hat A+}^{\mu_1\mu_2\mu_3}+\hat\Pi_{AA\hat A-}^{\mu_1\mu_2\mu_3},
\label{9.14}
\end{equation}
where
\begin{equation}
\begin{split}
\hat\Pi_{AA\hat A+}^{\mu_1\mu_2\mu_3}=&-e^{\frac{i}{2}p_1\theta p_2}\Big(\Pi_1^{\mu_1\mu_2\mu_3}\cdot I(p_1+p_2)+\Pi_2^{\mu_1\mu_2\mu_3}\cdot \hat I(p_1)+\Pi_3^{\mu_1\mu_2\mu_3}\cdot I_{+}
\\&+\Pi_4^{\mu_1\mu_2}(p_1,p_2)\cdot I^{\mu_3}_{+}+\Pi_4^{\mu_2\mu_3}(p_2,p_3)\cdot I^{\mu_1}_++\Pi_4^{\mu_1\mu_3}(p_1,p_3)\cdot I^{\mu_2}_+\Big),
\end{split}
\label{9.15}
\end{equation}
while
\begin{equation}
\hat\Pi_{AA\hat A-}^{\mu_1\mu_2\mu_3}=\hat\Pi_{AA\hat A+}^{\mu_1\mu_2\mu_3}\left(p_1\leftrightarrow p_2,\;\mu_1\leftrightarrow\mu_2\right).
\label{9.16}
\end{equation}
The above master integrals $\hat I(p_1)$, $I_+$ and $I_+^\mu$ bear the following forms:
\begin{gather}
\hat I(p_1)=\int\frac{d^D\ell}{(2\pi)^D}\frac{e^{-i\ell\theta(p_1+p_2)}}{\ell^2(\ell-p_1)^2},
\label{hI}\\
I_{+}=\int\frac{d^D\ell}{(2\pi)^D}\frac{e^{-i\ell\theta(p_1+p_2)}}{\ell^2(\ell-p_1)^2(\ell-p_1-p_2)^2},
\label{I+}\\
I^{\mu}_+=\int\frac{d^D\ell}{(2\pi)^D}\frac{\ell^\mu e^{-i\ell\theta(p_1+p_2)}}{\ell^2(\ell-p_1)^2(\ell-p_1-p_2)^2},
\label{Imu+}
\end{gather}
while the tensor structures are given below
\begin{gather}
\begin{split}
\Pi_1^{\mu_1\mu_2\mu_3}&=2\big(\eta^{\mu_2\mu_3}(p_1+p_2)^{\mu_1}-\eta^{\mu_1\mu_3}(p_1+p_2)^{\mu_2}\big),
\\
\Pi_2^{\mu_1\mu_2\mu_3}&=2\eta^{\mu_2\mu_3}p_2^{\mu_1},
\\
\Pi_3^{\mu_1\mu_2\mu_3}&=p_1^{\mu_1}(p_1+p_2)^{\mu_3}(2p_1+p_2)^{\mu_2}
+\eta^{\mu_1\mu_2}\big(p_1^{\mu_3}(2p_1\cdot p_2+p_2^2)-p_2^{\mu_3}p_1^2\big)
\\&
-\eta^{\mu_1\mu_3}\big(p_2^{\mu_2}p_1^2+p_1^{\mu_2}(2p_1\cdot(p_1+p_2)+p_2^2)\big)
+\eta^{\mu_2\mu_3}\big(p_2^{\mu_1}p_1^2-p_1^{\mu_1}(2p_1\cdot p_2+p_2^2)\big),
\\
\Pi_4^{\mu_1\mu_2}&=2(\eta^{\mu_1\mu_2}p_1\cdot p_2-p_1^{\mu_2}p_2^{\mu_1}).
\end{split}
\label{9.20}
\end{gather}
Now, by setting $\theta=0$ in the integrands of $\hat I(p_1)$, $I_{+}$ and $I^{\mu}_+$, one obtains integrals which are both UV finite and IR finite by power counting. Therefore one can apply Lebesque's dominated convergence theorem to conclude that the limit $\theta^{\mu\nu}\rightarrow 0$ of the tensor $\hat\Pi_{AA\hat A+}^{\mu_1\mu_2\mu_3}$ exists and is given by the corresponding Green function of the commutative ABJM field theory. It is plain that the analysis carried out for the $\hat\Pi_{AA\hat A+}^{\mu_1\mu_2\mu_3}$ tensor will apply to the tensor $\hat\Pi_{AA\hat A-}^{\mu_1\mu_2\mu_3}$ as well, so that the limit $\theta^{\mu\nu}\rightarrow 0$ of the latter is given by the corresponding Green function in the ordinary ABJM theory too. Putting it all together, one concludes that the limit $\theta^{\mu\nu}\rightarrow 0$ of the $\hat\Pi_{AA\hat A}^{\mu_1\mu_2\mu_3}$ tensor is given by the ordinary ABJM field theory.

We shall end this subsection by showing explicitly that $I{+}$ and $I^{\mu}_+$ have well defined limit when $\theta\to 0$. Both integrals $I{+}$ and $I^{\mu}_+$ can be evaluated using the standard Schwinger-Feynman parametrization, \cite{Boos:1987bg}. So, as an example, let us work out $I_{+}$:
\begin{gather}
\begin{split}
I_{+}=&\;i\int\limits_0^1 dx\,\int\limits_0^1 dy\,(1-y)\int\limits_0^\infty d\alpha\,\alpha^2
\int\frac{d^D\ell}{(2\pi)^D}\,e^{-\alpha\ell^2}
\cdot e^{ix(1-y)p_1\theta p_2}\cdot e^{-\alpha X-\frac{1}{4\alpha}(\tilde p_1+\tilde p_2)^2},
\end{split}
\label{I+1}
\end{gather}
where
\begin{equation}
X=(1-y)\Big(x(1-x)p_1^2-y\big(x(1-x)p_1^2-(1-x)(p_1+p_2)^2-xp_2^2\big)\Big).
\label{I+X}
\end{equation}
The integration over variables $\ell$ and $\alpha$ then yields Bessel K-functions:
\begin{gather}
\begin{split}
I_+=&\;i\;(4\pi)^{-\frac{D}{2}}\int\limits_0^1 dx\,\int\limits_0^1 dy(1-y)e^{ix(1-y)p_1\theta p_2}\cdot 2\cdot X^{\frac{D}{4}-\frac{3}{2}}
\\&\cdot\left(\frac{(\tilde p_1+\tilde p_2)^2}{4}\right)^{\frac{3}{2}-\frac{D}{4}}K_{3-\frac{D}{2}}\left[\sqrt{X(\tilde p_1+\tilde p_2)^2}\right].
\label{I+2}
\end{split}
\end{gather}
In order to analyze the commutative limit we rewrite the $D$-dimensional Bessel K-function as sum of two Bessel I-functions
\begin{equation}
\begin{split}
K_{3-\frac{D}{2}}\left[\sqrt{X(\tilde p_1+\tilde p_2)^2}\right]=&\frac{\pi}{2\sin\left(3-\frac{D}{2}\right)\pi}
\\&\cdot\left(I_{\frac{D}{2}-3}\left[\sqrt{X(\tilde p_1+\tilde p_2)^2}\right]-I_{3-\frac{D}{2}}\left[\sqrt{X(\tilde p_1+\tilde p_2)^2}\right]\right).
\end{split}
\label{BessI}
\end{equation}
The Bessel I-functions can then be expand as power series.\footnote{Unlike integral $I$, the integration over the Feynman parameters can only be performed in $D$-dimension here, therefore the expansion over $\theta$ is performed in $D$-dimension.}
Next we can observe that the power series with respect to $\theta$ converges for small $\theta$ and $D<4$, with the leading term matching the commutative scalar triangle in \cite{Boos:1987bg}. Therefore the commutative limit exists.

The integral $\hat I(p_1)$ may be estimated using the same method performed for the integral $I$ in the appendix C yielding  the following result
\begin{equation}
\begin{split}
\hat I(p_1)\Big|_{D\to 3}=&\frac{\sqrt{2}}{(4\pi)^{\frac{3}{2}}}\int\limits_0^1 dx\,e^{ix(p_1\theta p_2)}\left(\frac{x(1-x)p_1^2}{(\tilde p_1+\tilde p_2)^2}\right)^{-\frac{1}{4}}K_{\frac{1}{2}}\left[\sqrt{x(1-x)p_1^2(\tilde p_1+\tilde p_2)^2}\right]
\\
=&\frac{1}{4\pi}\int\limits_0^1 dx\,e^{ix(p_1\theta p_2)}\frac{e^{-\sqrt{x(1-x)p_1^2(\tilde p_1+\tilde p_2)^2}}}{\sqrt{x(1-x)p_1^2}}.
\end{split}
\label{hI+}
\end{equation}
For small $\theta$ power series expansion is regular and the commutative limit does exist.

In a view of the computations of the three point function $\big<A^{\mu_1} A^{\mu_2} \hat A^{\mu_3}\big>$ carried out above, it is apparent that the three point function $\big<\hat A^{\mu_1} \hat A^{\mu_2} A^{\mu_3}\big>$ also goes to the ordinary result when the noncommutative tensor $\theta^{\mu\nu}\to 0$.

\section{Scalar $|$ Fermion, $\big<X_A X^B\big>$ $|$ $\big<\Psi^A\bar\Psi_B\big>$,  two-point functions}
\begin{figure}[t]
\begin{center}
\includegraphics[width=10cm]{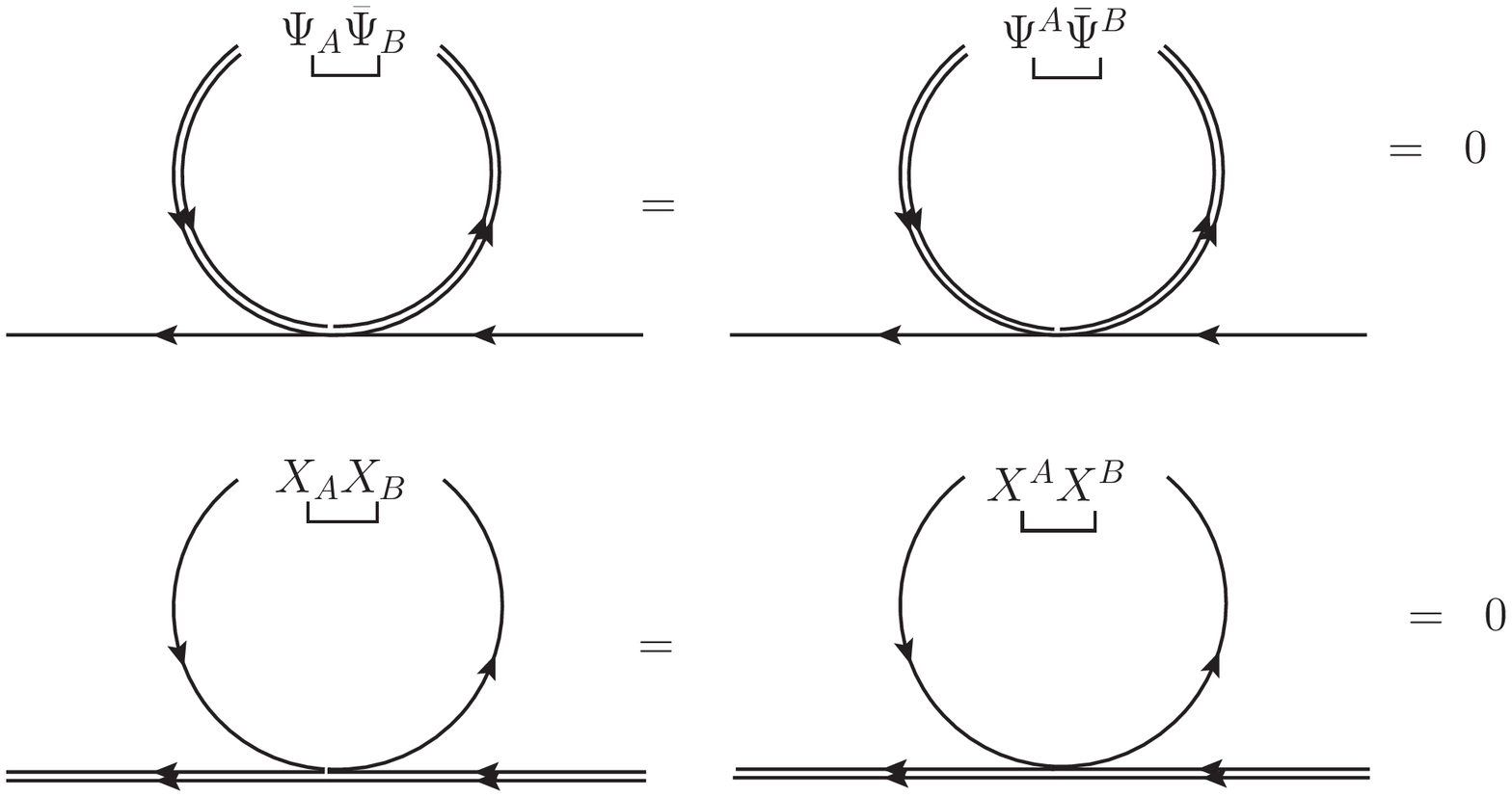}
\end{center}
\caption{Vanishing 1-loop contributions due to:
$\bcontraction{}{\Psi_A}{}{\bar\Psi_A}\Psi_A\bar\Psi_B=\bcontraction{}{\Psi^A}{}{\bar\Psi^B}\Psi^A\bar\Psi^B=\bcontraction{}{X_A}{} {X_B}X_A X_B=\bcontraction{}{X^A}{} {X^B}X^A X^B=0$.
}
\label{fig:Fig18N}
\end{figure}

From the four-field (2-scalars-2-fermions) action $S_4$   (\ref{AS4}) in accord with the given Feynman rules (3d diagram in figure \ref{fig:Fig25FR}, generically representing a number of diagrams as the one diagram), we have a number of contracted combinations of indices $A,B,C,D$. Since we have the following vanishing propagators: $\big<X_A X_B\big>= \big<X^A X^B\big>=\big<\Psi_A\bar\Psi_B\big>=\big<\Psi^A\bar\Psi^B\big>= 0$, the one-loop tadpole contributions to the 2-point functions coming from (\ref{AS4a}) part of the action  vanish. Namely as illustrated in figure \ref{fig:Fig18N}, we obtain vanishing contributions to the relevant tadpole diagrams due to the antisymmtric properties of Levi-Civita tensor $\epsilon^{ABCD}$ in (\ref{AS4a}), following filed contractions in the tadpole loops $\bcontraction{}{\Psi_A}{}{\bar\Psi_B}\Psi_A\bar\Psi_B$=$\bcontraction{}{\Psi^A}{}{\bar\Psi^B}\Psi^A\bar\Psi^B$=$\bcontraction{}{X_A}{} {X_B}X_A X_B$=$\bcontraction{}{X^A}{} {X^B}X^A X^B=0$.

To work out the one-loop contributions to $\big<X_A X^A\big>$ and $\big<\bar\Psi_A\Psi^A\big>$ we only need the vertices coming from (\ref{AS4b}) and (\ref{AS4c}) parts of the action (\ref{AS4}).

\subsection{One-loop Scalar $\big<X_AX^B\big>$ 2-point function}
\begin{figure}[t]
\begin{center}
\includegraphics[width=12cm]{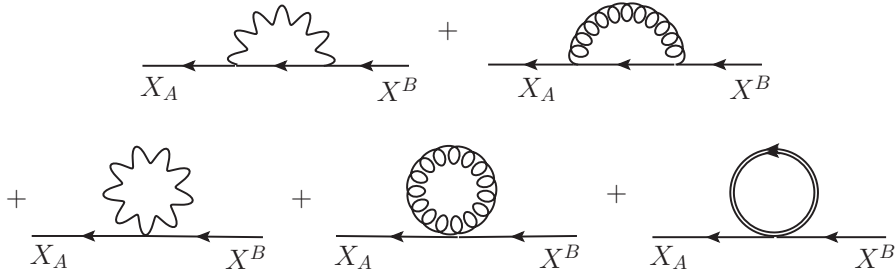}
\end{center}
\caption{Loop contributions to the scalar 2-point function $\big<X_A X^B\big>$.}
\label{fig:Fig19N}
\end{figure}
Using Feynman rules from appendix D, one can show that the integrand corresponding to the 1st Feynman diagram of figure \ref{fig:Fig19N} vanishes since the epsilon tensor of the gauge field propagator is contracted with two equal momenta. The 2nd diagram vanishes for the same reason. The integrands of the 3d and 4th diagrams are zero due to the contraction $\eta^{\mu\nu}\epsilon_{\mu\nu\rho}$ that occur in each of them. The last diagram --a digram absent in the ordinary theory-- also has a vanishing integrand since it carries factor (see the 3d Feynman rule (\ref{FRXXPsiPsi}) from figure \ref{fig:Fig25FR})
\begin{equation}
\sin\Big[\frac{1}{2}\big(p\theta p+\ell\theta\ell\big)\Big]\equiv0,
\label{10.1}
\end{equation}
where $p$ is the external momentum and $\ell$ is the loop momentum. Let us point out that $S_{4a}$ in (\ref{AS4}) does not contribute to the last diagram in figure \ref{fig:Fig19N}, since the free propagators
$\big<\Psi_A\bar\Psi_B\big>$ and $\big<\Psi^A\bar\Psi^B\big>$ vanish, respectively.

Putting it all together we conclude that the one-loop contribution to the scalar two-point function $\big<X_A X^B\big>$ (in the Landau gauge) vanishes in both, the noncommutative and the ordinary ABJM quantum field theories, respectively.

\subsection{One-loop Fermion $\big<\Psi^A\bar\Psi_B\big>$ 2-point function}

\begin{figure}[t]
\begin{center}
\includegraphics[width=12cm]{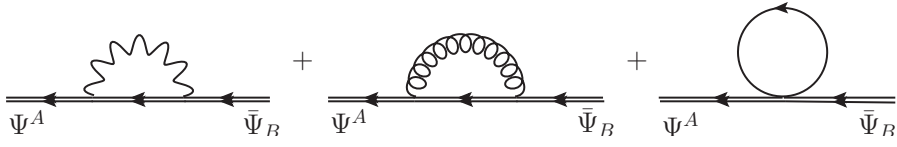}
\end{center}
\caption{Loop contributions to the fermion 2-point function $\big<\Psi^A\bar\Psi_B\big>$.}
\label{fig:Fig20N}
\end{figure}
Again using relevant Feynman rules from appendix D, one can show that the one-loop contribution to the $\big<\Psi^A \bar\Psi_B\big>$ (in the Landau gauge) vanishes in both, the noncommutative and the ordinary theories, respectively. Indeed, the integrands of the 1st two diagrams of figure \ref{fig:Fig20N} differ by a minus sign, so their sum vanishes. The integrand of the last diagram of figure \ref{fig:Fig20N} vanishes because it contains exactly the same vanishing factor as in eq. (\ref{10.1}). Let us point out that the action $S_{4a}$ in  (\ref{AS4}) does not contribute to the last diagram in figure \ref{fig:Fig20N}, since the free propagators $\big<X_A X_B\big>$ and $\big<X^A X^B\big>$ vanish --see i.e. figure \ref{fig:Fig18N}.

\section{Fermion -- gauge field $|$ -- hgauge field, $\big<\Psi^A\bar\Psi_B A^\mu\big>$ $|$ $\big<\Psi^A\bar\Psi_B \hat A^\mu\big>$, three-point functions}

\begin{figure}[t]
\begin{center}
\includegraphics[width=11cm]{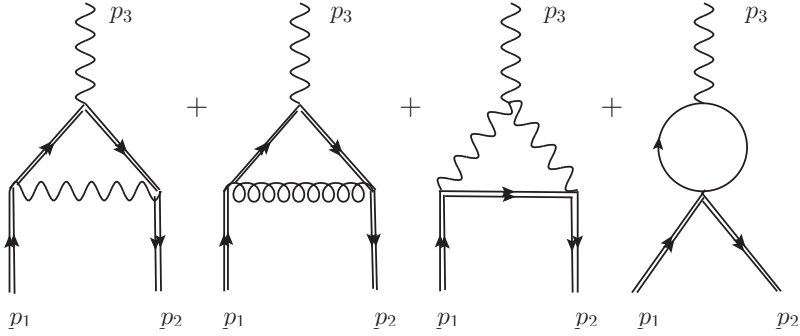}
\end{center}
\caption{Loop contributions to the 3-point functions $\big<\Psi^A(p_1) \bar\Psi_B(p_2) A^{\mu}(p_3)\big>$,  and $p_3=p_2-p_1$.}
\label{fig:Fig21}
\end{figure}

By using Feynman rules given in the appendix D it can be easily shown that the sum of the first two diagrams in figure \ref{fig:Fig21} reads
\begin{equation}
i\,\delta^A{}_B\;e^{\frac{i}{2}p_2\theta p_3}
\int \frac{d^D \ell}{(2\pi)^D} \Big[e^{i\ell\theta p_3} - 1\Big]\,
\frac{\big[\gamma^\nu (\slashed\ell+\slashed{ p}_2) \gamma^{\mu}(\slashed\ell+\slashed{ p}_1)\gamma^\rho\big]\epsilon_{\rho\nu\sigma}\ell^\sigma}{\ell^2(\ell+p_1)^2(\ell+p_2)^2},
\label{megustaone}
\end{equation}
where the incoming fermion has momentum $p_1$ and carries index A, the outgoing fermion has momentum $p_2$ and carries index B and the incoming gauge field has momentum $p_3=p_2-p_1$ and Lorentz index $\mu$.
Now, by expanding the integrand, the integral in (\ref{megustaone}) can be expressed as the following sum
\begin{equation}
\begin{array}{l}
{i\,\delta^A{}_B \;e^{\frac{i}{2}p_2\theta p_3}
\int \frac{d^D \ell}{(2\pi)^D} \Big[e^{i\ell\theta p_3} - 1\Big]\,
\dfrac{\big[\gamma^\nu\slashed\ell\gamma^{\mu}\slashed\ell\gamma^\rho\big]\epsilon_{\rho\nu\sigma}\ell^\sigma}{\ell^2(\ell+p_1)^2(\ell+p_2)^2}\;+}\\[4pt]
{i\,\delta^A{}_B \;e^{\frac{i}{2}p_2\theta p_3}
\int \frac{d^D \ell}{(2\pi)^D} \Big[e^{i\ell\theta p_3} - 1\Big]\,
\dfrac{\big[\gamma^\nu (\slashed\ell \gamma^{\mu}\slashed{ p}_1+\slashed{ p}_2\gamma^{\mu}\slashed\ell+\slashed{ p}_2\gamma^{\mu}\slashed{ p}_1)\gamma^\rho\big]\epsilon_{\rho\nu\sigma}\ell^\sigma}
{\ell^2(\ell+p_1)^2(\ell+p_2)^2}.
}
\end{array}
\label{megustatwo}
\end{equation}

Let us analyze the limit $\theta\to 0$ of the second intregral in (\ref{megustatwo}) at $D=3$. If we remove the factor
$e^{\frac{i}{2}p_2\theta p_3}\big[e^{i\ell\theta p_3} - 1\big]$
from the integral in question, we will end up with an integral that is UV finite and IR finite by power-counting for non-exceptional momenta. Hence, we can use Lebesgue's dominated convergence theorem and conclude that the limit $\theta\to 0$ of the second integral in (\ref{megustatwo}) can be computed by taking such limit under the integral sign; but this limit is zero. We have thus shown that in the limit  $\theta\to 0$ in the sum of the first two diagrams in figure \ref{fig:Fig21} only the first integral in (\ref{megustatwo}) contributes. After a little algebra and by using
$\gamma^{\nu}\gamma^{\rho}\gamma^{\mu}= \epsilon^{\nu\rho\mu}\mathbb{I}+\eta^{\rho\mu}\gamma^{\nu}-\eta^{\nu\mu}\gamma^{\rho}+\eta^{\nu\rho}\gamma^{\mu}$,
 one obtains
\begin{equation}
\begin{array}{l}
{i\,\delta^A{}_B \;e^{\frac{i}{2}p_2\theta p_3}
\int \frac{d^D \ell}{(2\pi)^D} \Big[e^{i\ell\theta p_3} - 1\Big]\,
\dfrac{\big[\gamma^\nu\slashed\ell\gamma^{\mu}\slashed\ell\gamma^\rho\big]\epsilon_{\rho\nu\sigma}\ell^\sigma}{\ell^2(\ell+p_1)^2(\ell+p_2)^2}=}\\[4pt]
{-2i\,\delta^A{}_B \;e^{\frac{i}{2}p_2\theta p_3}
\int \frac{d^D \ell}{(2\pi)^D} \Big[e^{i\ell\theta p_3} - 1\Big]\,
\dfrac{\ell^{\mu}}{(\ell+p_1)^2(\ell+p_2)^2}.
}
\end{array}
\label{megustathree}
\end{equation}

Let us now consider the sum of the last two diagrams in figure \ref{fig:Fig21}. Proceeding as above and  after some lengthy algebra, one concludes that, in the limit $\theta\to 0$, the sum of these two diagrams is given by
\begin{equation}
+2i\,\delta^A{}_B \;
\int \frac{d^D \ell}{(2\pi)^D} \big[e^{i\ell\theta p_3} e^{-\frac{i}{2}p_2\theta p_3}- e^{\frac{i}{2}p_2\theta p_3}\big]\,
\frac{\ell^{\mu}}{(\ell+p_1)^2(\ell+p_2)^2}.
\label{megustafour}
\end{equation}
Now, adding (\ref{megustathree}) and (\ref{megustafour}), for $D=3$ and $p_3=p_2-p_1$ we finally obtain
\begin{equation}
-2i\,\delta^A{}_B \;(e^{\frac{i}{2}p_2\theta p_3}- e^{-\frac{i}{2}p_2\theta p_3})\,
\int \frac{d^3 \ell}{(2\pi)^3}\,e^{i\ell\theta p_3}\,\frac{\ell^{\mu}}{(\ell+p_1)^2(\ell+p_2)^2},
\label{megustafive}
\end{equation}
which after changing of variables, $\ell\rightarrow -\ell-p_1$, gives
\begin{equation}
2i\,\delta^A{}_B \;(e^{\frac{i}{2}p_2\theta p_3}- e^{-\frac{i}{2}p_2\theta p_3})e^{-ip_1\theta p_3}\,
\int \frac{d^3 \ell}{(2\pi)^3}\,e^{-i\ell\theta p_3}\,\frac{\ell^{\mu}+p_1^{\mu}}{\ell^2(\ell-p_3)^2}.
\label{megustasix}
\end{equation}

Taking into account results presented in subsection C.2 of the appendix C, we conclude that the integral
\begin{equation}
\int \frac{d^3 \ell}{(2\pi)^3}\,e^{-i\ell\theta p_3}\,\frac{\ell^{\mu}+p_1^{\mu}}{\ell^2(\ell-p_3)^2},
\label{11.7}
\end{equation}
remains bounded --although its limit does not exist-- as $\theta^{\mu\nu}$ approaches zero. Hence, the vanishing $\theta^{\mu\nu}$ limit of the expression  in (\ref{megustasix}) is zero due to the vanishing factor
\begin{equation}
\big(e^{\frac{i}{2}p_2\theta p_3}- e^{-\frac{i}{2}p_2\theta p_3}\big)\Big|_{\theta\to 0}
=2i \sin\frac{p_2\theta p_3}{2}\bigg|_{\theta\to 0}=0.
\label{11.8}
\end{equation}

To summarize, we have shown that the limit $\theta^{\mu\nu}\rightarrow 0$ of the sum of all four diagrams in figure \ref{fig:Fig21} vanishes, being also UV finite for the nonvanishing $\theta^{\mu\nu}$.

Let us finally point out that in the ordinary ABJM field theory, with the gauge group being abelian, the last two diagrams in figure \ref{fig:Fig21} are absent,  besides the sum of the first two is zero. Indeed, this sum is obtained by setting $\theta=0$ in the expontetials in (\ref{megustaone}), i.e., by setting $\theta=0$ in the Feynman rules from the appendix D.

It is plain that the conclusion we have just reached for the one-loop 1PI contribution to $\big<\Psi^A\bar\Psi_B A^\mu\big>$ will also be valid for $\big<\Psi^A\bar\Psi_B\hat A^\mu\big>$, as a sum of contributions from the details of  figures \ref{fig:Fig21}, where the wavy gauge field lines are replaced with curly hgauge field lines and viceversa {\it (wavy $\leftrightarrow$ curly)}.

\section{Scalar -- gauge field  $|$ -- hgauge field, $\big<X^A X_B A^\mu\big>$ $|$  $\big<X^A X_B \hat A^\mu\big>$, three-point functions}

\begin{figure}[t]
\begin{center}
\includegraphics[width=12cm]{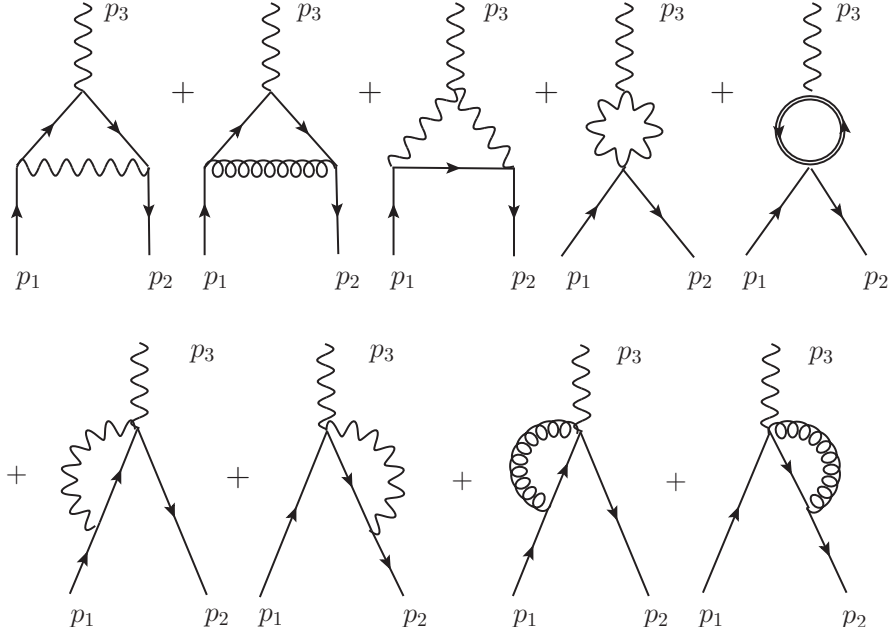}
\end{center}
\caption{Loop contributions to the 3-point functions $\big<X^A(p_1) X_B(p_2) A^\mu(p_3)\big>$, and $p_3=p_2-p_1$.}
\label{fig:Fig22}
\end{figure}

From Feynman rules in the appendix D we have the one-loop 1PI three-point function $\Gamma_{X^A X_B A^\mu}^\mu$ as a sum of contributions from the first and second line in figure \ref{fig:Fig22}, respectively
\begin{eqnarray}
\Gamma_{X^A X_B A^\mu}^\mu&=&S^{\mu}_{\rm tri1}+S^{\mu}_{\rm tri2}+
P^{\mu}_{\rm tri3}+P^{\mu}_{\rm bub1P}+F^{\mu}_{\rm bub2F}+
\nonumber\\
&+&S^{\mu}_{\rm legp1}+S^{\mu}_{\rm legp2}+
S^{\mu}_{\rm leghp1}+S^{\mu}_{\rm leghp2}.
\label{SSP}
\end{eqnarray}

Similarly we have the one-loop 1PI three-point function $\hat\Gamma_{X^A X_B \hat A^\mu}^\mu$ as a sum of contributions from the following detailed figure \ref{fig:Fig22} where the wavy gauge field lines are replaced with curly hgauge field lines and vice-versa {\it (wavy $\leftrightarrow$ curly)}
\begin{eqnarray}
\hat\Gamma_{X^A X_B \hat A^\mu}^\mu&=&
\hat S^{\mu}_{\rm tri1}+\hat S^{\mu}_{\rm tri2}+
\hat P^{\mu}_{\rm tri3}+\hat P^{\mu}_{\rm bub1P}+\hat F^{\mu}_{\rm bub2F}
\nonumber\\
&+&\hat S^{\mu}_{\rm leghp1}+\hat S^{\mu}_{\rm leghp2}+
\hat S^{\mu}_{\rm legp1}+ \hat S^{\mu}_{\rm legp2}.
\label{SShP}
\end{eqnarray}
Other contributions vanish due to the absence of relevant terms in the action. Remaining terms in (\ref{SSP}) and (\ref{SShP}) we compute below. 

We concentrate next on the $\Gamma_{X^A X_B A^\mu}^\mu$. The first three diagrams  listed in figure \ref{fig:Fig22}, i.e. the triangle diagrams, seem to be superficially logarithmic divergent without NC regulation. Explicit computation shows, however, that their divergence order are universally reduced by one because of the Levi-Civita tensor:
\begin{equation}
\begin{split}
S^{\mu}_{\rm tri1}=&i\delta^A{}_B\int\frac{d^D\ell}{(2\pi)^D}e^{-i\frac{p_1\theta p_2}{2}}e^{-i\ell\theta(p_1-p_2)}
\\&\cdot\frac{(2\ell+p_1-p_2)^\mu\epsilon_{\nu\rho\sigma}(\ell+2p_1-p_2)^\nu(\ell-p_2)^\rho(\ell+p_2)^\sigma}{\ell^2(\ell-p_2)^2(\ell+p_1-p_2)^2}
\\=&-4i\delta^A{}_B\int\frac{d^D\ell}{(2\pi)^D}e^{-i\frac{p_1\theta p_2}{2}}e^{-i\ell\theta(p_1-p_2)}\frac{(2\ell+p_1-p_2)^\mu\epsilon_{\nu\rho\sigma}\ell^\nu p_1^\rho p_2^\sigma}{\ell^2(\ell-p_2)^2(\ell+p_1-p_2)^2},
\end{split}
\label{12.3}
\end{equation}
similarly
\begin{equation}
\begin{split}
S^{\mu}_{\rm tri2}=4i\delta^A{}_B\int\frac{d^D\ell}{(2\pi)^D}e^{i\frac{p_1\theta p_2}{2}}\frac{(2\ell+p_1-p_2)^\mu\epsilon_{\nu\rho\sigma}\ell^\nu p_1^\rho p_2^\sigma}{\ell^2(\ell-p_2)^2(\ell+p_1-p_2)^2},
\end{split}
\label{12.4}
\end{equation}
and
\begin{equation}
\begin{split}
P^{\mu}_{\rm tri3}&=\delta^A{}_B\int\frac{d^D\ell}{(2\pi)^D}2
\frac{\epsilon^{\mu\nu\rho}\epsilon_{\delta\eta\nu}\epsilon_{\sigma\gamma\rho}(p_2-\ell-p_1)^{\eta}(p_1+p_2-\ell)^\delta(2p_2-\ell)^\sigma\ell^\gamma}{\ell^2(\ell-p_2)^2(\ell+p_1-p_2)^2}
\\&{\;\;\;\;}\cdot e^{i\frac{p_1\theta p_2}{2}}e^{\frac{i}{2}\ell\theta(p_1-p_2)}\sin\frac{\ell\theta(p_1-p_2)}{2}
\\&=-4\delta^A{}_B\int\frac{d^D\ell}{(2\pi)^D}e^{i\frac{p_1\theta p_2}{2}}\left(1-e^{-\ell\theta(p_1-p_2)}\right)\epsilon_{\nu\rho\sigma}\ell^\nu p_1^\rho p_2^\sigma(\ell^\mu-p_2^\mu).
\end{split}
\label{12.5}
\end{equation}
Lebesgue's dominated convergence theorem then rules these three integrals as continuous at the commutative limit. The remaining six bubble integrals are given below. The first two of them are symmetric under the exchange $p_1\to -p_2$.

Performing simple variable change $\ell\to -\ell+p_1-p_2$, we found the following expression for the 4th diagram in figure \ref{fig:Fig22}:
\begin{equation}
\begin{split}
P^{\mu}_{\rm bub1P}&=-\delta^A{}_B \int\frac{d^D\ell}{(2\pi)^D} 2\sin\frac{\ell\theta (p_1-p_2)}{2}e^{\frac{i}{2}p_1\theta p_2}\Big(e^{\frac{i}{2}\ell\theta(p_1-p_2)}+e^{-\frac{i}{2}\ell\theta(p_1-p_2)}\Big)
\\&\;\;\;\;\cdot\frac{\epsilon^{\rho\sigma\mu}\epsilon\indices{_{\rho\delta}^\eta}\epsilon_{\eta\gamma\sigma}\ell^\delta(\ell-p_1+p_2)^\gamma}{\ell^2(\ell-p_1+p_2)^2}
\\&=-\delta^A{}_B \int\frac{d^D\ell}{(2\pi)^D}2\sin\ell\theta (p_1-p_2)e^{\frac{i}{2}p_1\theta p_2}\frac{\epsilon\indices{_{\rho\sigma}^\mu}(p_1-p_2)^\rho\ell^\sigma}{\ell^2(\ell-p_1+p_2)^2}
\\&=-2i\delta^A{}_B \int\frac{d^D\ell}{(2\pi)^D}e^{\frac{i}{2}p_1\theta p_2}e^{-i\ell\theta(p_1-p_2)}\frac{\epsilon\indices{_{\rho\sigma}^\mu}(p_1-p_2)^\rho\ell^\sigma}{\ell^2(\ell-p_1+p_2)^2},
\end{split}
\label{12.6}
\end{equation}
while for the 5th diagram in figure \ref{fig:Fig22} we have:
\begin{equation}
\begin{split}
F^{\mu}_{\rm bub2F}=\delta^A{}_B \int\frac{d^D\ell}{(2\pi)^D}2e^{-\frac{i}{2}\ell\theta(p_1-p_2)}(4-2)\sin\frac{p_1\theta p_2+\ell\theta(p_1-p_2)}{2}\frac{\tr(\slashed\ell-\slashed p_1+\slashed p_2)\gamma^\mu\slashed\ell}{\ell^2(\ell-p_1+p_2)^2}.
\label{12.7}
\end{split}
\end{equation}
Using the fact $\tr\gamma^\mu\gamma^\nu\gamma^\rho=2\epsilon^{\mu\nu\rho}$, we conclude that the above contribution is
\begin{equation}
\begin{split}
F^{\mu}_{\rm bub2F}=&8\delta^A{}_B \int\frac{d^D\ell}{(2\pi)^D}e^{-\frac{i}{2}\ell\theta(p_1-p_2)}\sin\frac{p_1\theta p_2+\ell\theta(p_1-p_2)}{2}\frac{\epsilon\indices{_{\rho\sigma}^\mu}(p_1-p_2)^\rho\ell^\sigma}{\ell^2(\ell-p_1+p_2)^2}
\\=&4i\delta^A{}_B \;\int\frac{d^D\ell}{(2\pi)^D}\Big(e^{-\frac{i}{2}p_1\theta p_2}e^{-i\ell\theta(p_1-p_2)}-e^{\frac{i}{2}p_1\theta p_2}\big)\frac{\epsilon\indices{_{\rho\sigma}^\mu}(p_1-p_2)^\rho\ell^\sigma}{\ell^2(\ell-p_1+p_2)^2}.
\end{split}
\label{12.8}
\end{equation}
Next four asymmetric bubble diagrams from figure \ref{fig:Fig22} are as follows:
\begin{equation}
S^{\mu}_{\rm legp1}=-\delta^A{}_B \int\frac{d^D\ell}{(2\pi)^D}2ie^{\frac{i}{2}p_1\theta p_2}\Big(1+e^{-i\ell\theta(p_1-p_2)}\Big)\frac{\epsilon\indices{_{\rho\sigma}^\mu}p_1^\rho\ell^\sigma}{\ell^2(\ell+p_1)^2},
\label{12.9}
\end{equation}
\begin{equation}
S^{\mu}_{\rm legp2}=-\delta^A{}_B \int\frac{d^D\ell}{(2\pi)^D}2ie^{\frac{i}{2}p_1\theta p_2}\Big(1+e^{i\ell\theta(p_1-p_2)}\Big)\frac{\epsilon\indices{_{\rho\sigma}^\mu}p_2^\rho\ell^\sigma}{\ell^2(\ell-p_2)^2},
\label{12.10}
\end{equation}
\begin{equation}
S^{\mu}_{\rm leghp1}=\delta^A{}_B \int\frac{d^D\ell}{(2\pi)^D}4ie^{\frac{i}{2}p_1\theta p_2}\frac{\epsilon\indices{_{\rho\sigma}^\mu}p_1^\rho\ell^\sigma}{\ell^2(\ell+p_1)^2},
\label{12.11}
\end{equation}
\begin{equation}
S^{\mu}_{\rm leghp2}=\delta^A{}_B \int\frac{d^D\ell}{(2\pi)^D}4ie^{\frac{i}{2}p_1\theta p_2}\frac{\epsilon\indices{_{\rho\sigma}^\mu}p_2^\rho\ell^\sigma}{\ell^2(\ell-p_2)^2}.
\label{12.12}
\end{equation}
Then, it is not hard to see that integrals $S^{\mu}_{\rm leghp1}$ and $S^{\mu}_{\rm leghp2}$ are planar, while the nonplanar part of the remaining integrals involve the same master integral $\hat I^\mu(p,\theta q)$ which is evaluated in the appendix C.2, with a common $q=p_1-p_2$ up to the $\pm$ sign. Furthermore, the Levi-Civita symbols suppress all $p^\mu$ terms in nonplanar integrals as well as all planar integrals.

Finally we are left with the following sum of the leading order terms from (\ref{SSP})
\begin{equation}
\begin{split}
\Gamma_{X^A X_B A^\mu}^\mu\sim & \frac{i}{8\pi}\,\epsilon\indices{_{\rho\sigma}^\mu}\Big(\big(-2(p_1-p_2)^\rho-2p_1^\rho+2p_2^\rho\big)e^{i\frac{p_1\theta p_2}{2}}
\\&+4(p_1-p_2)^\rho e^{-i\frac{p_1\theta p_2}{2}}\Big)\frac{(\tilde p_1-\tilde p_2)^\sigma}{\sqrt{(\tilde p_1-\tilde p_2)^2}}
\\=&\frac{\epsilon\indices{_{\rho\sigma}^\mu}(p_1-p_2)^\rho}{\pi}\frac{(\tilde p_1-\tilde p_2)^\sigma}{|\tilde p_1-\tilde p_2|}\sin\frac{p_1\theta p_2}{2},
\end{split}
\label{12.13}
\end{equation}
which clearly vanishes when $\theta\to 0$. This concludes our discussion on the existence of the commutative limit result and its equivalence to the corresponding result obtained by working within ordinary ABJM quantum field theory. The latter is obtained by setting $\theta^{\mu\nu}=0$ in the integrands of each integral above, prior to the integration over the loop momentum.

In view of the computations carried out, it is plain that the limit $\theta^{\mu\nu}\to 0$ of the 1PI contribution to the 3-point function $\big<X^A X_B \hat A^\mu\big>$ exists and matches the ordinary result.

\section{Summary and Discussion}

In this paper we have formulated a quantum ABJM field theory on the noncommutative spacetime as defined by the Moyal star-product. By using component formalism we have shown that the theory has an ${\cal N}=6$ supersymmetry. We have done so by defining the supersymmetry transformations of the noncommutative fields which generalize the ordinary ones and leave the noncommutative classical action invariant. Next, we have considered the noncommutative ${\rm U}_{\kappa}(1)\times {\rm U}_{-\kappa}(1)$ field theory --this theory is radically different from its ordinary counterpart since it is nonabelian-- and we have analyzed the existence of the noncommutativity matrix $\theta^{\mu\nu}\to 0$ limit of each one-loop 1PI function with fewer --barring ghosts-- than four fields.  We have shown that this limit exists and it is given by the corresponding Green function of the ordinary ABJM quantum field theory, a result which only trivially holds for all one-loop UV convergent (by power counting) 1PI Green functions.
Along the way we have found out that the computed Green functions turned out to be not UV divergent, although they were not UV finite by power counting either. Of course, we have also seen that power counting and Lebesgue's dominated convergence theorem immediately lead to the conclusion that, if the UV degree of divergence is negative, the limit $\theta^{\mu\nu}\to 0$ of the one-loop 1PI functions is given by the ordinary ABJM quantum field theory results  --see section 4, for further details.

As far as our computations can tell the noncommutative ABJM field theory does not contain any noncommutative IR divergence and, therefore, it has no noncommutative IR instability. The noncommutative ABJM quantum field theory put forward here makes an excellent candidate for well defined noncommutauve gauge field theory which turns into the ordinary ABJM quantum field theory as the noncommutativity matrix $\theta^{\mu\nu}$ approaches to zero.

Putting it all together, we can conclude that we have introduced a consistent noncommutative deformation of the ordinary ABJM quantum field theory, this being a chief asset of the paper. Of course, many properties of the noncommutative theory remain to be studied. One most essential task in the authors' minds is to carry out checks which could verify that indeed the noncommutative quantum field theory of the $\rm U(N)_{\kappa}\times U(N)_{-\kappa}$ generalization of our construction will be the gauge dual of the deformed noncommutative gravity theory constructed in \cite{Imeroni:2008cr}.

In this article we have shown that our construction does possess the same $\mathcal N=6$ supersymmetry and, by construction shares the same multiplet as the undeformed theory. Therefore the next check which must be carried out is to match the correlation functions by using the standard prescription:
\begin{equation}
\Big< {\rm exp}\Big[\int\!d^3k\,\phi_0(k){\cal O}(k)\Big]\Big> = e^{-S_{\rm SUGRA}\big(\phi(k,u)\big)},
\label{12.1}
\end{equation}
where $\phi_0(k)$ is the boundary value (in Fourier space) of the bulk field $\phi(k,u)$, and ${\cal O}(k)$ denotes generically  the nonlocal composite operators in \cite{Gross:2000ba}. See \cite{Maldacena:1999mh} and references therein, for further details. On the left hand side of equation (\ref{12.1}) occur the correlation functions of the nonlocal composite operators, ${\cal O}(k)$, discussed in \cite{Gross:2000ba}, which are to be computed in the noncommutative ABJM quantum field theory. The values and properties of the 1PI functions studied in this paper is definitely one solid step towards elucidating the properties and computing the values of the correlation functions on the left hand side of the equation (\ref{12.1}). One has to, however, bear in mind some unique difficulties in this program: The first and foremost one from the authors' viewpoint comes from one crucial basic property of the (Moyal type) noncommutative deformation, which could be called {\it planar equivalence rule}~\cite{Filk:1996dm,Douglas:2001ba}: This {\it rule} states that because of the multiplication consistency relations \eqref{Multcon}, the planar diagrams of the noncommutative field theory, in the sense of the (star product analogy to the) color ordering~\cite{Bern:1990ux} sense\footnote{A color decomposition is convenient to show properties like for example figures 7, 8 and 9 are nonplanar. Yet we did not use it in the computation presented in this article as they are simple enough without it. Color decomposition can be very beneficial for more complicated amplitudes in NCABJM theory for sure.}, contain no loop momenta dependent NC phase factor and therefore remain the same as in the commutative theory from the loop integral perspective. One the other hand the most successful developments in the {\em undeformed} gauge/gravity duality program are inherently in the planar limit. This makes a direct comparison uneasy as the planar limit on the field theory side misses the unique NC features in the quantum corrections as we have seen above\footnote{On the other hand, we have also seen that the planar diagrams still carry the NC phase factors depending on the external momenta. It has been worked out in the dipole deformation that the phase factor structure, as it is, can be nontrivial for crucial subject(s) in gauge/gravity duality like integrability~\cite{Guica:2017mtd}. We thank Jun-bao Wu for pointing this out to us.}, while the information in the
nonplanar amplitudes could be uneasy to obtain from the dual gravity/string theory side. Also, the dual gravity backgrounds constructed for both $\mathcal N=4$ NCSYM and NCABJM shares the same property that the NC directions of the metric become degenerate at the $r\to\infty$ boundary, which could raise quite subtle questions in the holographic correlation function computation~\cite{Landsteiner:2007bd}. While to carry out the checks that validate (\ref{12.1}) lies outside the scope of this paper, we would like to stress that the most decisive check should be on the correspondence between {\em nonplanar} amplitudes obtained from both field theory and dual gravity/string theory sides. We would be absolutely delighted if some new checks of the gauge/gravity duality with noncommutative deformation at the amplitude level can be done in near future.

Other issues that should be addressed are whether there is a supersymmetry enhancement at levels $\kappa= 1,2$, and, of course,  whether the results presented in this paper regarding UV finiteness and the limit $\theta^{\mu\nu}\rightarrow 0$ hold at any order in perturbation theory and for the $\rm U(N)$ gauge groups.

Finally, it would be very interesting to apply nonperturbative methods \cite{Marino:2011nm, Perez:2013dra} to the noncommutative quantum ABJM field theory introduced in this paper.

\section{Acknowledgements}
The work by C.P. Martin has been financially supported in part by the Spanish MINECO through grant FPA2014-54154-P. This work is also supported by the Croatian Science Foundation (HRZZ) under Contract No. IP-2014-09-9582, and
we acknowledge the support of the COST Action MP1405  (QSPACE). J. You acknowledges support by the H2020 Twining project No. 692194, RBI-T-WINNING, and would like to acknowledge the support of W.~Hollik and the Max-Planck-Institute for Physics, Munich, for hospitality. We also thank Johanna Erdmenger, Karl Landsteiner and Jun-bao Wu for many discussions on gauge/gravity duality and/or ABJM theory.

\appendix

\section{$\rm SU(4)_R$ supersymmetric invariance of $\rm U(1)_\kappa\times U(1)_{-\kappa}$ theory}

In the following analysis all the volume integrals are like in the action being 3-dimensional. We integrate over $d^3x$, and denoted it as the integral only, i.e. the notation is $\int d^3x \equiv \int$.

\subsection{Variations of the action with respect to gauge and scalar fields}

For the noncommutative Chern--Simons term SUSY transformation $\delta$ reads
\begin{equation}
\delta S_{\rm CS} =\frac{\kappa}{2\pi} \int  \, \epsilon^{\mu\nu\rho}
\frac{1}{2}\big(\delta A_\mu\star F_{\nu\rho}-\delta \hat A_\mu\star \hat F_{\nu\rho}\big),
\label{deltaCS}
\end{equation}
while for the first scalar field kinetic term from (\ref{Akin}) we have found
\begin{eqnarray}
\delta S_{\rm kinS} &=& \frac{-\kappa}{2\pi} \int
 \bigg[D^\mu \delta X^A\star D_\mu X_A
+\big(D^\mu X^A\big)\star D_\mu\delta X_A
\nonumber\\
&+&\Big(i\delta\hat A^\mu\star X^A-iX^A\star\delta A^\mu\Big)D_\mu X_A
+\big(D^\mu X^A\big)\Big(i\delta A_\mu\star X_A-iX_A\star\delta \hat A_\mu\Big)\bigg],
\nonumber\\
&=&\delta_1 S_{\rm kinS}+\delta_2 S_{\rm kinS}+\overline{\delta_1S_{\rm kinS}}+
\overline{\delta_2S_{\rm kinS}},
\label{2.1}\\
\delta_1 S_{\rm kinS} &=& \frac{\kappa}{2\pi} \int \big(D^2X^A\big)\star \delta X_A,
\label{2.2}\\
\overline{\delta_1S_{\rm kinS}} &=&\big(\delta_1 S_{\rm kinS}\big)^{*}=
\frac{\kappa}{2\pi} \int  \delta X^A\star D^2 X_A,
\label{2.3}\\
\delta_2 S_{\rm kinS} &=& \frac{-i\kappa}{2\pi} \int
\Big(\delta A_\mu X_A\star D^\mu X^A-\delta \hat A_\mu \big(D^\mu X^A\big)\star X_A\Big),
\nonumber\\
\overline{\delta_2 S_{\rm kinS}} &=& \frac{-i\kappa}{2\pi} \int
\Big(\delta \hat A^\mu X^A\star  D_\mu X_A-\delta A^\mu \big(D_\mu X_A\big)\star X^A \Big),
\label{deltaXS}
\end{eqnarray}
and for the second fermionic kinetic term we finally have
\begin{eqnarray}
\delta S_{\rm kinF} &=& \frac{\kappa}{2\pi} \int
 \Big(i\delta\bar\Psi_A \star\slashed{D} \Psi^A+i\bar\Psi_A \star\slashed{D} \delta\Psi^A- \bar\Psi_A\star \delta\slashed{A} \Psi^A+\bar\Psi_A\star\gamma^\mu \Psi^A\delta\hat{A_\mu}\Big),
\nonumber\\
&=&
\delta_1 S_{\rm kinF}+\overline{\delta_1S_{\rm kinF}}+\delta_2 S_{\rm kinF},
\label{3.1}\\
\delta_1 S_{\rm kinF} &=& \frac{i\kappa}{2\pi} \int  \delta\bar\Psi_A\star\slashed{D}\Psi^A ,
\label{3.2}\\
\overline{\delta_1 S_{\rm kinF}} &=& \frac{-i\kappa}{2\pi} \int \big(D^\mu \bar\Psi_A\big)\star\gamma_\mu\delta\Psi^A ,
\label{3.3}\\
\delta_2 S_{\rm kinF} &=& \frac{-\kappa}{2\pi} \int
\Big(\bar\Psi_A\star\delta\slashed{A} \Psi^A -\bar\Psi_A\gamma^\mu \star \Psi^A \delta\hat A_\mu\Big).
\label{3.4}
\end{eqnarray}

Now from (\ref{deltaCS},\ref{2.1},\ref{2.2},\ref{2.3},\ref{3.2},\ref{3.3}) with help of (\ref{CovD},\ref{SUSYtransf}) and $\epsilon^{\mu\nu\rho}\gamma_\mu=\gamma^{\nu\rho}$ we have
\begin{eqnarray}
\delta S_{\rm CS} &=& \frac{\kappa}{2\pi} \int  \,
\frac{1}{2}\bigg(\Gamma^I_{AB} \bar\epsilon^I\gamma^{\mu\nu} \Psi^A\star X^B-\tilde\Gamma^{IAB} X_B\star\bar\Psi_A\gamma^{\mu\nu}\epsilon^I  \bigg)
\star F_{\mu\nu}
\nonumber\\
&-&
\frac{\kappa}{2\pi} \int  \,
\frac{1}{2}\bigg(\Gamma^I_{AB}X^B\star\bar\epsilon^{I}\gamma^{\mu\nu} \Psi^A
-\tilde\Gamma^{IAB} \bar\Psi_A\gamma^{\mu\nu}\epsilon^I\star X_B\bigg)
\star \hat F_{\mu\nu},
\label{deltaCS1}
\end{eqnarray}
\begin{equation}
\delta_1 S_{\rm kinS}+\overline{\delta_1S_{\rm kinS}}
= \frac{i\kappa}{2\pi} \int
\bigg(\big(D^2X^A\big)\star\Gamma^I_{AB} \bar\epsilon^I  \Psi^B
-\tilde\Gamma^{IAB} \bar\Psi_B\epsilon^I \star D^2 X_A \bigg),
\label{1+1S}
\end{equation}
\begin{eqnarray}
\delta_1 S_{\rm kinF}+\overline{\delta_1S_{\rm kinF}}
&=&\frac{\kappa}{2\pi} \int \, \bigg[
-i\big(D^2X^A\big) \Gamma^I_{AB} \bar\epsilon^I\star \Psi^B
+i\tilde\Gamma^{IAB} \bar\Psi_B\star\epsilon^I D^2 X_A
\nonumber\\
&&\phantom{XXX}-\frac{1}{2}\Big(\Gamma^I_{AB} \bar\epsilon^{I}\gamma^{\mu\nu}\Psi^A\star X^B   -\tilde\Gamma^{IAB} X_B\star\bar\Psi_A\gamma^{\mu\nu}\epsilon^I \Big)\star F_{\mu\nu}
\nonumber\\
&&\phantom{XXX}+\frac{1}{2}\Big(\Gamma^I_{AB}X^B\star \bar\epsilon^{I}\gamma^{\mu\nu} \Psi^A
-\tilde\Gamma^{IAB} \bar\Psi_A\gamma^{\mu\nu}\epsilon^I\star X_B\Big)\star \hat F_{\mu\nu}
\nonumber\\
&&\phantom{XXXx}+iN^I_A\bar\epsilon^I\star\slashed{D}\Psi^A-i\big(D_\mu\bar\Psi_A\big)\star \gamma^\mu N^{IA}\epsilon^I
\bigg],
\label{1+1F}
\end{eqnarray}
which gives:
\begin{eqnarray}
\delta S_{\rm CS} +\delta_1 S_{\rm kinS}+\overline{\delta_1S_{\rm kinS}}+\delta_1 S_{\rm kinF}+\overline{\delta_1S_{\rm kinF}}&=&
\frac{i\kappa}{2\pi} \int  \,
\Big(N^I_A\bar\epsilon^I\star\slashed{D}\Psi^A-\big(D_\mu\bar\Psi_A\big)\star \gamma^\mu N^{IA}\epsilon^I\Big).
\nonumber\\
\label{1+1SF}
\end{eqnarray}
Finally we obtain:
\begin{eqnarray}
&&\hspace{-1cm}\delta S_{\rm CS}+\delta S_{\rm kinS}+\delta S_{\rm kinF}
=
\delta S_{\rm CS} +\delta_1 S_{\rm kinS}+\overline{\delta_1S_{\rm kinS}}
+\delta_1 S_{\rm kinF}+\overline{\delta_1S_{\rm kinF}}
\nonumber\\
&&\hspace{-.8cm}+
\frac{\kappa}{2\pi} \int  \, \bigg[
-\Big(i\delta\hat A^\mu\star X^A-iX^A\star\delta A^\mu\Big)D_\mu X_A
-\big(D^\mu X^A\big)\Big(i\delta A_\mu\star X_A-iX_A\star\delta \hat A^\mu\Big)
\nonumber\\
&&\phantom{XXxxx}\hspace{-.7cm}-\bar\Psi_A\star\delta \slashed{A}\Psi^A+\bar\Psi_A\star\gamma^\mu \Psi^A\delta\hat A_\mu+
iN^I_A\bar\epsilon^I\star\slashed{D}\Psi^A-i\big(D_\mu\bar\Psi_A\big)\star N^{IA}\gamma^\mu\epsilon^I\bigg].
\label{1+1SFfinal}
\end{eqnarray} 

\subsection{Variations of the action with respect to fermion fields}

Let us first define two variations with respect to fermion fields as a sum
\begin{equation}
\delta\Psi=\delta_1\Psi + \delta_3\Psi.
\label{Vdelta13Psi}
\end{equation}
where both variations acting on fermion fields give,  respectively
\begin{eqnarray}
\delta_1\Psi_A&=&\;\;\:\Gamma^I_{AB}\gamma^\mu \epsilon^I\star D_\mu X^B,
\hspace{.5cm}
\:\,\delta_3\Psi_A=N^I_A\star\epsilon^I,
\nonumber\\
\delta_1\bar\Psi_A&=&-\Gamma^I_{AB} \bar\epsilon^I\star\slashed{D}  X^B,
\hspace{1.1cm}
\;\,\delta_3\bar\Psi_A=N^I_A\star\bar\epsilon^I,
\nonumber\\
\delta_1\Psi^A&=&-\tilde\Gamma^{IAB} \gamma^\mu\epsilon^I\star D_\mu X_B,
\hspace{.57cm}
\delta_3\Psi^A=N^{IA}\star\epsilon^I,
\nonumber\\
\delta_1\bar\Psi^A&=&\;\;\:\tilde\Gamma^{IAB} \bar\epsilon^I\star\slashed{D} X_B,
\hspace{1.2cm}
\delta_3\bar\Psi^A=N^{IA}\star\bar\epsilon^I.
\label{delta13Psi}
\end{eqnarray}
Now we find a variation of the action $S_4$ with respect to the variation $\delta_1\Psi$:
\begin{eqnarray}
\delta_{\delta_1\Psi}S_4&=&\frac{i\kappa}{2\pi} \int  \,\bigg[ \tilde\Gamma^{IBC}
\bigg(2\bar\Psi_A\star\gamma^\mu\epsilon^I\star D_\mu(X_B\star X^A\star X_C)
\label{deltadelta1S4}\\
&+&\bar\Psi_B\star\gamma^\mu\epsilon^I\star
\Big(2X_C\star \big(D_\mu X^A\big)\star X_A-2X_A\star \big(D_\mu X^A\big)\star X_C
\nonumber\\
&+&\big(D_\mu X_C\big)\star X^A\star X_A-X_A\star X^A\star D_\mu X_C\bigg)
\nonumber\\
&-&2\epsilon_{ABCD}\delta_1\bar\Psi^A\star X^B\star\Psi^C\star X^D
+\bar\Psi^A\star\delta_1\Psi_A\star X_B\star X^B
\nonumber\\
&-&\delta_1\bar\Psi_A\star\Psi^A\star X^B\star X_B
+2\delta_1\bar\Psi_A\star\Psi^B\star X^A\star X_B-2\bar\Psi^B\star\delta_1\Psi_A\star X_B\star X^A\bigg].
\nonumber
\end{eqnarray}
Second, performing the variation with respect to the gauge fields
in the kinetic terms of the $X$'s and $N^{IA}\star\epsilon^I$ we have found
\begin{eqnarray}
\delta_{\delta A,\delta\hat A} S_{\rm kinS} &=& \frac{-i\kappa}{2\pi} \int
 \bigg[\Big(\delta\hat A^\mu\star X^A-X^A\star\delta A^\mu\Big)\star D_\mu X_A
+\big(D^\mu X^A\big)\star\Big(\delta A_\mu\star X_A-X_A\star\delta \hat A_\mu\Big)\bigg]
\nonumber\\
&=&
 \frac{i\kappa}{2\pi} \int \bigg[\tilde\Gamma^{IBC}\bar\Psi_B\star\gamma^\mu
 \epsilon^I\star\Big(X_C\star X^A\star D_\mu X_A-\big(D_\mu X_A\big)\star X^A\star X_C
 \nonumber\\
 &&\phantom{XXXXXXXXXX}+X_A\star \big(D^\mu X^A\big)\star X_C-X_C\star \big(D_\mu X^A\big)\star X_A\Big)
\nonumber\\
&&
\phantom{XXx} -\Gamma^I_{BC}\bar\epsilon^I\star\gamma^\mu\Psi^B\star
 \Big(X^A\star \big(D_\mu X_A\big)\star X^C-X^C\star \big(D_\mu X_A\big)\star X^A
 \nonumber\\
 &&\phantom{XXXXXXXXXX}+X^C\star X_A\star D_\mu X^A-\big(D^\mu X^A\big)\star X_A\star X^C\Big)
 \bigg].
 \label{deltaAdeltahatA}
\end{eqnarray}
Next after summing (\ref{deltadelta1S4}) and (\ref{deltaAdeltahatA}) we have total contribution as
\begin{eqnarray}
\delta_{\delta_1\Psi}S_4&+&\delta_{\delta A,\delta\hat A}S_{\rm kinS}=\frac{i\kappa}{2\pi} \int  \,\bigg[ \tilde\Gamma^{IBC}
\Big(2\bar\Psi_A\star\gamma^\mu\epsilon^I\star D_\mu\big(X_B\star X^A\star X_C\big)
\nonumber\\
&+&\bar\Psi_B\star\gamma^\mu\epsilon^I\star
D_\mu \big(X_C\star  X^A\star X_A-X_A\star  X^A\star X_C\big)\Big)
\nonumber\\
&-&2\epsilon_{ABCD}\delta_1\bar\Psi^A\star X^B\star\Psi^C\star X^D
+\bar\Psi^A\star\delta_1\Psi_A\star X_B\star X^B
\nonumber\\
&-&\delta_1\bar\Psi_A\star\Psi^A\star X^B\star X_B
+2\delta_1\bar\Psi_A\star\Psi^B\star X^A\star X_B-2\bar\Psi^B\star\delta_1\Psi_A\star X_B\star X^A
\nonumber\\
&-&\Gamma^I_{BC}\bar\epsilon^I\star\gamma^\mu\Psi^B\star
 \Big(X^A\star \big(D_\mu X_A\big)\star X^C-X^C\star \big(D_\mu X_A\big)\star X^A
 \nonumber\\
 &&\phantom{XXXXXXXXXX}+X^C\star X_A\star D_\mu X^A-\big(D^\mu X^A\big)\star X_A\star X^C\Big) \bigg],
 \label{delta1S4deltaA}
\end{eqnarray}
which should cancel against the variation of $S_{\rm kinF}$ induced by
$\delta_3\Psi$.

To prove the above statement lets first perform $\delta_3$ variation
\begin{eqnarray}
\delta_{\delta_3\Psi} S_{\rm kinF}&=&\frac{i\kappa}{2\pi} \int
 \Big(\delta_3\bar\Psi_A \star\slashed{D} \Psi^A+\bar\Psi_A \star\slashed{D} \delta_3\Psi^A\Big)
\nonumber\\
&=& \frac{i\kappa}{2\pi} \int \bar\Psi_A \star\slashed{D}N^{IA}\epsilon^I +C.C.
=\frac{i\kappa}{2\pi} \int \bar\Psi_A \star\gamma^\mu\epsilon^I D_\mu N^{IA} +C.C.
\label{D3PsikinF}
\end{eqnarray}
After summing a number of terms from (\ref{delta1S4deltaA}) and (\ref{D3PsikinF}) we have our prof verified, i.e.
\begin{equation}
\delta_{\delta_1\Psi}S_4+\delta_{\delta A,\delta\hat A}S_{\rm kinS}+\delta_{\delta_3\Psi} S_{\rm kinF}=0, \;\; \;\;\rm Q.E.D.
\label{proof2}
\end{equation}

\subsection{Cancellations between $S_{\rm CS}$, $S_{kin}$ and $S_4$ variations}

Let $\Psi_1,\Psi_2,\chi_3$ be spinors, then the integral
\begin{equation}
\int \Big(\Psi_{1i}\star\bar\Psi_2\star\chi_3+\Psi_{2i}\star\bar\chi_3\star\Psi_1+\chi_{3i}\star\bar\Psi_1\star\Psi_2\Big)=0, \;\forall i=1,2.
\label{spinors123}
\end{equation}
In our actual computations either of the spinor above may be a $\star$-product of one of our $\Psi$-spinor and one of our scalars $X$, i.e.  $\chi_3=\Psi_3\star X$.

Next we present the simplification to the following contribution from (\ref{1+1SFfinal}):
\begin{eqnarray}
&&\frac{\kappa}{2\pi} \int
\Big(-\bar\Psi_A\star\delta \slashed{A}\Psi^A+\bar\Psi_A\star\gamma^\mu \Psi^A\delta\hat A_\mu\Big)
\nonumber\\
&&
\phantom{XXXXXX}=\frac{\kappa}{2\pi} \int
\Big[i\bar\Psi^A\star\Psi_A\star\delta X_C\star X^C-i\bar\Psi_A\star\Psi^A\star X^C\star \delta X_C
\label{PsiAPsi}\\
&&\phantom{XXXXXXXXXXXX}+2\Gamma^I_{BC}\bar\epsilon^I\Psi^A\star
\big(\bar\Psi_A\star\Psi^B\star X^C-X^C\star\bar\Psi^B\star\Psi_A\big)+C.C.\Big].
\nonumber
\end{eqnarray}

Now we compute the variation of the 2nd term of $S_4$ induced by 
$\delta{X^B}$ and obtain:
\begin{eqnarray}
&&\delta_{X^B}\Big[\frac{-i\kappa}{2\pi}\epsilon_{ABCD}\int \bar\Psi^A \star X^B \star\Psi^C \star X^D \Big]
\nonumber\\
&&\phantom{XXXXX}=\frac{\kappa}{2\pi}\int
\Big[-2i\bar\Psi^A\star\Psi_A\star\delta X_B\star X^B+2i\bar\Psi_A\star\Psi^A \star X^B\star\delta X_B
\label{2ndS4}\\
&&\phantom{XXXXXXXXXXX}
+2i\bar\Psi^A\star\Psi_B\star\delta X_A\star X^B-2i\bar\Psi_A\star\Psi^B\star X^A\star\delta X_B
\nonumber\\
&&\phantom{XXXXXXXXXXXX}
-2\Gamma^I_{BC}\bar\epsilon^I\Psi^A
\star\big(\bar\Psi_A\star\Psi^B\star X^C-X^C\star\bar\Psi^B\star\Psi_A\big)\Big].
\nonumber
\end{eqnarray}
To work it out we have to use the cyclicality of the $\star$-product, i.e. employ
\begin{equation}
\int \Big(\bar\epsilon^I\star\Psi^E\star\Psi_\kappa^C+\Psi_{E\kappa}\star\bar\epsilon^I\star\Psi^C+\epsilon_\kappa^I\star\bar\Psi_E\star\Psi^C\Big)=0, \;\forall \kappa=1,2.
\label{cicly}
\end{equation}
Adding up (\ref{2ndS4}) and (\ref{PsiAPsi}) minus $C.C.$ part we have found
\begin{eqnarray}
&&\delta_{X^B}\Big[\frac{-i\kappa}{2\pi}\epsilon_{ABCD}\int  \bar\Psi^A \star X^B \star\Psi^C \star X^D \Big]
\nonumber\\
&& 
\phantom{XXXXXX}+\frac{\kappa}{2\pi} \int
\Big[i\bar\Psi^A\star\Psi_A\star\delta X_C\star X^C-i\bar\Psi_A\star\Psi^A\star\delta X^C\star X_C
\nonumber\\
&&\phantom{XXXXXXXXXXXX}+2\Gamma^I_{BC}\bar\epsilon^I\Psi^A\star
\big(\bar\Psi_A\star\Psi^B\star X^C-X^C\star\bar\Psi^B\star\Psi_A\big)\Big]
\nonumber\\
&&\phantom{XXXXX}=\frac{i\kappa}{2\pi}\int
\Big[-\bar\Psi^A\star\Psi_A\star\delta X_B\star X^B+\bar\Psi_A\star\Psi^A\star X^B\star\delta X_B
\nonumber\\
&&\phantom{XXXXXXXXXX}
+2\bar\Psi^A\star\Psi_B\star\delta X_A\star X^B -2\bar\Psi_A\star\Psi^B\star X^A\star\delta X_B
\Big].
\label{Addingup1}
\end{eqnarray}
Next we show that (\ref{Addingup1}) cancels against the variations of sum of the 3rd, 4th, 5th and 6th terms of $S_4$ induced by $\delta X_B$. After some computations those variations give:
\begin{equation}
\frac{i\kappa}{2\pi}\int
\Big[\bar\Psi^A\star\Psi_A\star\delta X_B\star X^B-\bar\Psi_A\star\Psi^A\star X^B\star\delta X_B
+2\bar\Psi_A\star\Psi^B\star X^A\star\delta X_B-2\bar\Psi^B\star\Psi_A\star \delta X_B\star X^A \Big],
\label{Addingup2}\\
\end{equation}
and it does cancel exactly above expression (\ref{Addingup1}),   Q.E.D.

Now we show that variations of $C.C.$ terms in (\ref{PsiAPsi}) cancels against the variation $\delta S_4$ which is a sum of the $\delta X_A$ variation of the first term in $S_4$ and the $\delta X^A$ variation of the 3rd, 4th, 5th and 6th terms of $S_4$
\begin{eqnarray}
\delta_{X_A}&\equiv&\delta_{X_A}\Big[\frac{\kappa}{2\pi}\int i\epsilon^{ABCD} \bar\Psi_A\star  X_B\star \Psi_C\star  X_D \Big]
\nonumber\\
&=&\frac{\kappa}{2\pi}\int
\Big[2i\bar\Psi_A\star\Psi^A\star\delta X^B\star X_B-2i\bar\Psi^A\star\Psi_A\star X_B\star\delta X^B
-2i\bar\Psi_A\star\Psi^B\star\delta X^A\star X_B
\nonumber\\
&&\phantom{X}+2i\bar\Psi^A\star\Psi_B\star X_A\star \delta X^B
-2\tilde\Gamma^{IBC}\bar\epsilon^I\star\Psi_A\star
\big(\bar\Psi^A\star\Psi_B\star X_C-X_C\star\bar\Psi_B\star\Psi^A\big)\Big],
\nonumber\\
\label{C.C.}
\end{eqnarray}
\begin{eqnarray}
\delta_{X^A}&\equiv&\delta_{X^A}\Big[\frac{i\kappa}{2\pi}\int
\bar\Psi^A\star\Psi_A\star X_B\star X^B-\bar\Psi_A\star\Psi^A \star X^B \star X_B
\nonumber\\
&&\phantom{XX}
+2\bar\Psi_A\star\Psi^B\star X^A\star X_B-2\bar\Psi^B\star\Psi_A\star X_B\star X^A
\Big]
\nonumber\\
&=&\frac{i\kappa}{2\pi}\int \Big[
\bar\Psi^A\star\Psi_A\star X_B\star\delta X^B-\bar\Psi_A\star\Psi^A\star \delta X^B\star X_B
\nonumber\\
&&\phantom{XX}
+2\bar\Psi_A\star\Psi^B\star \delta X^A\star X_B-2\bar\Psi^B\star\Psi_A\star X_B\star\delta X^A\Big],
\label{VarA3456}
\end{eqnarray}
\begin{eqnarray}
\delta S_4&=&\delta_{X_A}+\delta_{X^A}=
\frac{\kappa}{2\pi}\int
\Big[i\bar\Psi_A\star\Psi^A\star\delta X^B\star X_B-i\bar\Psi_A\star\Psi^A\star X_B\star\delta X^B
\nonumber\\
&&\phantom{XXXXXXXX}-2\tilde\Gamma^{IBC}\bar\epsilon^I\star\Psi_A\star
\big(\bar\Psi^A\star\Psi_B\star X_C-X_C\star\bar\Psi_B\star\Psi^A\big)\Big].
\label{C.C.S4}
\end{eqnarray}
Finally we denote the $C.C.$ terms from eq. (\ref{PsiAPsi}) as $\delta_{C.C.}(\ref{PsiAPsi})$ and obtain:
\begin{eqnarray}
\delta_{C.C.} (\ref{PsiAPsi})&=&
\frac{\kappa}{2\pi}\int
\Big[i\bar\Psi^A\star\Psi_A\star X_B\star\delta X^B-i\bar\Psi_A\star\Psi^A\star\delta X^B\star X_B
\nonumber\\
&&\phantom{XX}+2\tilde\Gamma^{IBC}\bar\epsilon^I\star\Psi_A\star
\big(\bar\Psi^A\star\Psi_B\star X_C-X_C\star\bar\Psi_B\star\Psi^A\big)\Big],
\label{C.C.B2}
\end{eqnarray}
which shows perfect match, i.e. the full cancelation as expected:
\begin{equation}
\delta_{C.C.} (\ref{PsiAPsi})+\delta S_4=0, \;\;\,\;Q.E.D.
\label{proof3}
\end{equation}

\subsection{Classical supersymmetric invariance regarding $S_6$ terms}

The last step to show the full SUSY invariance of the noncommutative ABJM action is to confirm that the $\delta_3$ transformation of the fermions in the $\Psi^2 X^2$ terms of the action $S_4$ is cancelled by the $\delta=\delta_{X_A}+\delta_{X^A}$ transformation of $X^3\star X^3\sim X^6$ order terms in the action $S_6$. As already given before, the $X^3$ order transformation needed bears the form 
\begin{equation}
\delta_3\Psi^A= N^{IA}\star\epsilon^I,\,\delta_3\Psi_A=N^I_A\star\epsilon^I,\,\delta_3\bar\Psi^A=N^{IA}\star\bar\epsilon^I,\,\delta_3\bar\Psi_A=N^I_A\star\bar\epsilon^I,
\label{C1}
\end{equation}
where
\begin{gather}
N^{IA}=\tilde\Gamma^{IAB}\left(X_C\star X^C\star X_B-X_B\star X^C\star X_C\right)-2\tilde\Gamma^{IBC}X_B\star X^A\star X_C,
\label{A.34}\\
N^I_A=\Gamma^{I}_{AB}\left(X^C\star X_C\star X^B-X^B\star X_C\star X^C\right)-2\Gamma^I_{BC}X^B\star X_A\star X^C.
\label{C2}
\end{gather}
Notice that $\Psi$'s are Majorana fermions, i.e. $\bar\Psi=\Psi^T\gamma^0$, therefore the variation of $\Psi$ and $\bar\Psi$ can be identified as the same if they carry the identical index.

Taking into account the Majorana nature of the fermions, the overall cyclicality under the star/matrix product, matrix trace and integration, as well as the definition of fermion contraction, we have found the following $\delta_3$ variation of the action $S_4$
\begin{equation}
\delta_3 S_4=\Delta_1+\Delta_2,
\label{C3}
\end{equation}
\begin{equation}
\begin{split}
\Delta_1&=i\int-2\epsilon_{ABCD}\;{\rm tr}\,\Big(\delta_3\bar\Psi^A\star X^B\star \Psi^C\star X^D\Big)
\\&+{\rm tr}\;\delta_3\bar\Psi_A\star\Big(X_B\star X^B\star \Psi^A-\Psi^A\star X^B\star X_B+2\Psi^B\star X^A\star X_B-2X_B\star X^A\star \Psi^B\Big),
\end{split}
\label{S4variation1}
\end{equation}
\begin{equation}
\begin{split}
\Delta_2&=i\int 2\epsilon^{ABCD}\;{\rm tr}\,\Big(\delta_3\bar\Psi_A\star X_B\star X_C\star X_D\Big)
\\&+{\rm tr}\;\delta_3\bar\Psi^A\star\Big(\Psi_A\star X_B\star X^B-X^B\star X_B\star\Psi_A+2X^B\star X_A\star \Psi_B-\Psi_B\star X_A\star X^B\Big).
\end{split}
\label{C5}
\end{equation}
Since $\Delta_1$ and $\Delta_2$ can be handled in practically identical way, we concentrate on the first one only. Substituting definitions of $\delta_3\bar\Psi^A$ and $\delta_3\bar\Psi_A$ we have
\begin{equation}
\begin{split}
\Delta_1=&i\int-\epsilon_{ABCD}\epsilon^{AEFG}\Gamma^I_{FG}\bar\epsilon^I{\rm tr}\Big(X_H\star X^H\star X_E\star X^B\star\Psi^C\star X^D
\\&-X_E\star X^H\star X_H\star X^B\star\Psi^C\star X^D\Big)
\\&
+ 2\epsilon_{ABCD}\epsilon^{EFGH}\Gamma^I_{GH}\bar\epsilon^I{\rm tr}
X_E\star X^A\star X_F\star X^B\star\Psi^C\star X^D
\\&
+{\rm tr}\Big(\Gamma^I_{AD}\big(X^C\star X_C\star X^D-X^D\star X_C\star X^C\big)
-2\Gamma^I_{CD}X^C\star X_A\star X^D\Big)
\\&\cdot
\bar\epsilon^I\Big(X_B\star X^B\star\Psi^A
-\Psi^A\star X^B\star X_B+2\Psi^B\star X_A\star X_B-2X_B\star X^A\star\Psi^B\Big),
\end{split}
\label{C6}
\end{equation}
where we used the identity $2\tilde\Gamma^{IAB}=\epsilon^{ABCD}\Gamma^I_{CD}$.
Next we recall two fundamental identities of the Levi-Civita symbols
\begin{equation}
\epsilon_{ABCD}\epsilon^{AEFG}=\delta_{BCD}^{EFG}, 
\;\;\;\;
\epsilon_{ABCD}\epsilon^{EFGH}=\delta_{ABCD}^{EFGH},
\label{C7}
\end{equation}
where the generalized Kronecker $\delta$-symbol is defined as follows
\begin{equation}
\delta^{j_1...j_n}_{i_1...i_n}=\sum_{\sigma\in S_4}{\rm sign}(\sigma)\delta_{i_1}^{j_{\sigma(1)}}......\delta_{i_n}^{j_{\sigma(n)}}.
\label{C8}
\end{equation}
After employment of the generalized Kronecker $\delta$-symbols and some lengthy yet straightforward algebra, the transformation \eqref{S4variation1} boils down to
\begin{equation}
\begin{split}
\Delta_1=&i\Gamma^I_{AB}\bar\epsilon^I\tr\int -\Psi^B\star\Big(X^A\star X_C\star X^C\star X_D\star X^D
\\&+X^C\star X_C\star X^D\star X_D\star X^A+4X^C\star X_D\star X^A\star X_C\star X^D\Big)
\\&+2\Psi^B\star \Big(X^A\star X_C\star X^D\star X_D\star X^C
\\&+X^C\star X_D\star X^D\star X_C\star X^A+X^C\star X_C\star X^A\star X_D\star X^D\Big).
\end{split}
\label{C9}
\end{equation}
We can then easily recognize that all $\Psi^A$ 's in the formula above are contracted with $\Gamma^I_{AB}$ from the $X_A$ supersymmetric transformation in (\ref{SUSYtransf}): $\delta X_A=i\Gamma^I_{AB}\bar\epsilon^I\star\Psi^B$. All other terms cancel each other, thus we can rewrite it as
\begin{equation}
\begin{split}
\Delta_1=&-\frac{1}{3}\delta_{X_A}\tr\int\; X_A\star X^A\star X_B\star X^B\star X_C\star X^C
+X^A\star X_A\star X^B\star X_B\star X^C\star X_C
\\&+4X_A\star X^B\star X_C\star X^A\star X_B\star X^C
-6X_A\star X^B\star X_B\star X^A\star X_C\star X^C,
\end{split}
\label{C11}
\end{equation}
and this is exactly $\Delta_1=-\delta_{X_A}S_6$. Similarly one can show that $\Delta_2=-\delta_{X^A}S_6$, therefore
\begin{equation}
\delta_3 S_4=-\delta S_6, \;\;\;\rm Q.E.D.
\label{C12}
\end{equation}

\section{Two-point functions: $\big<A^\mu A^\nu\big>$}

We notice that due to the bi-fundamental nature of the fermions and scalar bosons, they do not form non-planar contribution to the one-loop identical gauge field amplitudes. This fact reduces the relevant diagrams for identical gauge fields to pure gauge field theory (gauge field and ghost loops) only. We use the following convention for the one-loop purely gauge field diagrams:\\
$\bullet$ For each vertex, assignment of momenta is in such sequence: outgoing loop momenta, external momenta, incoming loop momenta.\\
$\bullet$ Each propagator has the 1st index as outgoing, and the 2nd index as incoming (the 3d index on of the $\epsilon$-tensor is contracted with the momentum flowing through the propagator).\\
As an example let's write down the gauge field bubble $P^{\mu\nu}_{\rm bub}$, figure \ref{fig:Figl2}, in this convention

\begin{equation}
\begin{split}
P^{\mu\nu}_{\rm bub}=&\frac{1}{2}\int\frac{d^D\ell}{(2\pi)^D}(-2i)^2\sin\frac{\ell\theta(p-\ell)}{2}\epsilon^{\mu_2\mu\mu_1}
\frac{\epsilon_{\mu_1\nu_1\rho_1}\ell^{\rho_1}}{\ell^2}
\sin\frac{(p-\ell)\theta\ell}{2}\epsilon^{\nu_1\nu\nu_2}
\frac{\epsilon_{\nu_2\mu_2\rho_2}(\ell-p)^{\rho_2}}{(\ell-p)^2}
\\
=&\int\frac{d^D\ell}{(2\pi)^D}2\sin^2\frac{\ell\theta p}{2}\frac{\epsilon^{\mu_2\mu\mu_1}\epsilon_{\mu_1\rho_1\nu_1}\ell^{\rho_1}\epsilon^{\nu_1\nu\nu_2}\epsilon_{\nu_2\rho_2\mu_2}(\ell-p)^{\rho_2}}{\ell^2(\ell-p)^2}.
\end{split}
\label{B.1}
\end{equation}
We then evaluate the contraction of Levi-Civita symbols in three dimensions as guided by the dimensional reduction convention, which yields
\begin{equation}
P^{\mu\nu}_{\rm bub}=\int\frac{d^D\ell}{(2\pi)^D}2\sin^2\frac{\ell\theta p}{2}\frac{\ell^
\mu(\ell-p)^\nu+\ell^\nu(\ell-p)^\mu}{\ell^2(\ell-p)^2}.
\label{B.2}
\end{equation}
We can then load one more transformation $\ell\to -\ell+p$ to turn the first half of the result above to be identical to the second half and obtain
\begin{equation}
P^{\mu\nu}_{\rm bub}=\int\frac{d^D\ell}{(2\pi)^D}4\sin^2\frac{\ell\theta p}{2}\frac{(\ell-p)^\mu\ell^\nu}{\ell^2(\ell-p)^2}.
\label{B.3}
\end{equation}

Next let us turn to the ghost bubble $G^{\mu\nu}_{\rm bub}$, figure \ref{fig:Figl3},
\begin{equation}
\begin{split}
G^{\mu\nu}_{\rm bub}=&\int\frac{d^D\ell}{(2\pi)^D}(-)^3(-2i)^2\frac{(\ell-p)^\mu\ell^\nu}{\ell^2(\ell-p)^2}\sin\frac{p\theta\ell}{2}\sin\frac{-p\theta(\ell-p)}{2}
\\=&-\int\frac{d^D\ell}{(2\pi)^D}4\sin^2\frac{\ell\theta p}{2}\frac{(\ell-p)^\mu\ell^\nu}{\ell^2(\ell-p)^2}.
\end{split}
\label{B.4}
\end{equation}
Thus
\begin{equation}
P^{\mu\nu}_{\rm bub}+G^{\mu\nu}_{\rm bub}=0,
\label{B.5}
\end{equation}
i.e. all potentially non-planar contributions cancel out.

\section{Integrals from two-point functions}

During this work we studied new integrals and found some new relations among them. Here we present a set of seven integrals $I,I_1$,....,$I_6$ appearing in (\ref{ShPbub}) relevant to this work. They are used to present all loop integral results in the main text. We start with D-dimensions and for the Euclidian signature:
\begin{eqnarray}
I&=&\frac{i}{(4\pi)^{D/2}}
\int_0^1 dx\int_0^\infty d\lambda \;\lambda^{1-D/2}\;
e^{-\lambda p^2x(1-x) -\frac{\tilde p^2}{4\lambda}},
\label{C.1}\\
I_1&=&\frac{i}{2(4\pi)^{D/2}}
\int_0^1 dx\int_0^\infty d\lambda \;\lambda^{-D/2}\;
e^{-\lambda p^2x(1-x) -\frac{\tilde p^2}{4\lambda}}=iI_6,
\label{C.2}\\
I_2&=&\frac{i}{(4\pi)^{D/2}}
\int_0^1 dx \;x^2\int_0^\infty d\lambda\; \lambda^{1-D/2}\;
e^{-\lambda p^2x(1-x) -\frac{\tilde p^2}{4\lambda}},
\label{C.3}\\
I_3&=&\frac{1}{2(4\pi)^{D/2}}
\int_0^1 dx \;x\int_0^\infty d\lambda \;\lambda^{-D/2}\;
e^{-\lambda p^2x(1-x) -\frac{\tilde p^2}{4\lambda}},
\label{C.4}\\
I_4&=&\frac{-i}{4(4\pi)^{D/2}}
\int_0^1 dx\int_0^\infty d\lambda \;\lambda^{-1-D/2}\;
e^{-\lambda p^2x(1-x) -\frac{\tilde p^2}{4\lambda}},
\label{C.5}\\
I_5&=&\frac{i}{2(4\pi)^{D/2}}
\int_0^1 dx \;x\int_0^\infty d\lambda \;\lambda^{1-D/2}\;
e^{-\lambda p^2x(1-x) -\frac{\tilde p^2}{4\lambda}},
\label{C.6}
\end{eqnarray}
where we recall that
$\tilde p^\mu=\theta^{\mu\nu}p_\nu$ and $\tilde p^0=\theta^{0\nu}p_\nu=0$.
After using
\begin{equation}
\int_0^\infty d\lambda \;\lambda^{-\nu}\;
e^{- p^2x(1-x)\lambda -\frac{\tilde p^2}{4\lambda}}=
2^\nu\Big(\frac{p^2x(1-x)}{\tilde p^2}\Big)^{\frac{\nu-1}{2}}
K_{\nu-1}\Big(\sqrt{\tilde p^2 p^2x(1-x)}\Big),
\label{C.7}
\end{equation}
and specifying $D=3$, integrals reduce to
\begin{eqnarray}
I&=&\frac{i}{2(2\pi)^{3/2}}
\int_0^1 dx \Big(\frac{p^2x(1-x)}{\tilde p^2}\Big)^{-1/4}
K_{-1/2}\Big(\sqrt{\tilde p^2 p^2x(1-x)}\Big).
\label{C.8}\\
I_1&=&\frac{i}{2(2\pi)^{3/2}}
\int_0^1 dx \Big(\frac{p^2x(1-x)}{\tilde p^2}\Big)^{1/4}
K_{1/2}\Big(\sqrt{\tilde p^2 p^2x(1-x)}\Big),
\label{C.9}\\
I_2&=&\frac{i}{2(2\pi)^{3/2}}
\int_0^1 dx \;x^2\Big(\frac{p^2x(1-x)}{\tilde p^2}\Big)^{-1/4}
K_{-1/2}\Big(\sqrt{\tilde p^2 p^2x(1-x)}\Big),
\label{C.10}\\
I_3&=&\frac{1}{2(2\pi)^{3/2}}
\int_0^1 dx \;x\Big(\frac{p^2x(1-x)}{\tilde p^2}\Big)^{1/4}
K_{1/2}\Big(\sqrt{\tilde p^2 p^2x(1-x)}\Big),
\label{C.11}\\
I_4&=&\frac{-i\sqrt{2}}{(2\pi)^{3/2}}
\int_0^1 dx \Big(\frac{p^2x(1-x)}{\tilde p^2}\Big)^{3/4}
K_{3/2}\Big(\sqrt{\tilde p^2 p^2x(1-x)}\Big),
\label{C.12}\\
I_5&=&\frac{i}{2(2\pi)^{3/2}}
\int_0^1 dx \;x\Big(\frac{p^2x(1-x)}{\tilde p^2}\Big)^{-1/4}
K_{-1/2}\Big(\sqrt{\tilde p^2 p^2x(1-x)}\Big),
\label{C.13}
\end{eqnarray}

\subsection{Master scalar integral $I(p,\theta)$}

Integral $I(p,\theta)$ is UV finite when $D<4$, therefore polarization tensor $\hat\Pi_{A\hat A}^{\mu\nu}$ has a smooth commutative limit. To verify this we employ the standard Schwinger-Feynman parametrization, which yields:
\begin{equation}
\begin{split}
I(p,\theta)\Big|_{D\to 3}=&\frac{\sqrt{2}}{(4\pi)^{\frac{3}{2}}}\int\limits_0^1 dx\,\left(\frac{x(1-x)p^2}{\tilde p^2}\right)^{-\frac{1}{4}}K_{\frac{1}{2}}\left[\sqrt{x(1-x)p^2\tilde p^2}\right]
\\
=&\frac{1}{4\pi}\int\limits_0^1 dx\,\frac{e^{-\sqrt{x(1-x)p^2\tilde p^2}}}{\sqrt{x(1-x)p^2}}.
\end{split}
\label{C.14}
\end{equation}
To get back to the Minkowski signature of our integral expressions (\ref{ShPbub}), (\ref{ShPtad}), and (\ref{FhPbub}) we apply simple transformations of say $(k,p)$ pair of momenta: $k^0\to-ik^0$, and $p^0\to-ip^0 \Longrightarrow p^2\to p^2-i0^+$, and then under the Wick rotations, performed by making a change on the righthand side of  our integrals (\ref{C.1}-\ref{C.6}), we obtain:
\begin{equation}
(I^M,I^M_i)= (I,I_i)\Bigg|_{\hspace{-3mm}{\substack
{k^0\to-ik^0\\p^0\to-ip^0\\ \hspace{4mm}p^2\to p^2-i0^+}}},\,\;\forall i=1,....,6.
\label{C.15}
\end{equation}
There are two relations among above integrals which makes results (\ref{ShPbub})--(\ref{FhPbub}) simpler:
\begin{eqnarray}
2I^M_3-I^M_6&=&0,
\label{C.16}\\
4\Big(I^M_2-I^M_5\Big)+I^M&=&
\bigg(-4I^M_1+\frac{i}{2\pi\sqrt{\tilde p^2}} \bigg)\frac{1}{(p^2-i0^+)}.
\label{C.17}
\end{eqnarray}

\subsection{An integral with a bounded but ill-defined $\tilde p^\mu\rightarrow 0$ limit}

Let us analyze the limit $\tilde p^\mu\rightarrow 0$ of the following integral:
\begin{equation}
I^\mu(p,\tilde p)=\int \frac{d^3 \ell}{(2\pi)^3}
e^{-i\ell\theta p}\;\frac{\ell^\mu}{\ell^2(\ell-p)^2},
\label{C.18}
\end{equation}
which is, for large loop momenta, the dominant contribution to  the diagram in figure \ref{fig:Figl15} --see (\ref{Stribub3}). By introducing Schwinger parameters we decompose integral (\ref{C.18}) into
\begin{equation}
I^\mu(p,\tilde p)= I_5\,p^\mu\,+\,I_6\,\tilde p^\mu,
\label{C.19}
\end{equation}
where integrals $I_5$ and $I_6$ have been defined in (\ref{C.2}), (\ref{C.6}) and (\ref{C.13}), respectively. Taking into account that
\begin{equation}
K_{\pm \frac{1}{2}}(z)=\sqrt{\frac{\pi}{2}}\,\frac{e^{-z}}{\sqrt{z}},
\label{C.20}
\end{equation}
one can show that integral (\ref{C.18}) further boils down to
\begin{equation}
I^\mu(p,\tilde p)= \frac{1}{8\pi}\,\frac{\tilde p^\mu}{\sqrt{\tilde p^2}}+\frac{i}{16}\frac{ p^\mu}{\sqrt{ p^2}}\,+\, f^\mu(p,\tilde p),
\label{C.21}
\end{equation}
where $f^\mu(p,\tilde p)$ vanishes as $\tilde p^\mu\rightarrow 0$.

Notice that the first summand on the right hand side of (\ref{C.21}) is bounded as $\tilde p^\mu\rightarrow 0$, but this limit depends on the way one approaches $\tilde p^\mu = 0$ point. To conclude, the limit $\tilde p^\mu\rightarrow 0$ of integral $I^\mu(p,\tilde p)$ (\ref{C.18}) is ill-defined, though not divergent.

The discussion above can be generalized to the following $D$-dimensional integral
\begin{equation}
\hat I^\mu(p,\tilde q)=\int \frac{d^3 \ell}{(2\pi)^3}
e^{-i\ell\theta q}\;\frac{\ell^\mu}{\ell^2(\ell-p)^2}=\hat I_5\,p^\mu\,+\,\hat I_6\,\tilde q^\mu,
\label{C.22}
\end{equation}
with
\begin{gather}
\begin{split}
\hat I_5=&\frac{i}{2(4\pi)^{D/2}}
\int_0^1 dx \;xe^{-ixp\theta q}\int_0^\infty d\lambda \;\lambda^{1-D/2}\;
e^{-\lambda p^2x(1-x) -\frac{\tilde q^2}{4\lambda}},
\end{split}
\label{C.23}
\\
\begin{split}
\hat I_6=&\frac{1}{2(4\pi)^{D/2}}
\int_0^1 dx \;e^{-ixp\theta q}\int_0^\infty d\lambda \;\lambda^{-D/2}\;
e^{-\lambda p^2x(1-x) -\frac{\tilde q^2}{4\lambda}}.
\end{split}
\label{C.24}
\end{gather}
When setting $D=3$, integrals $\hat I_5$ and $\hat I_6$ boils down to the following forms
\begin{gather}
\hat I_5=\frac{i}{2(2\pi)^{3/2}}
\int_0^1 dx \;xe^{-ixp\theta q}\Big(\frac{p^2x(1-x)}{\tilde q^2}\Big)^{-1/4}
K_{-1/2}\Big(\sqrt{\tilde q^2 p^2x(1-x)}\Big),
\label{C.25}
\\
\hat I_6=\frac{1}{2(2\pi)^{3/2}}
\int_0^1 dx e^{-ixp\theta q}\Big(\frac{p^2x(1-x)}{\tilde q^2}\Big)^{1/4}
K_{1/2}\Big(\sqrt{\tilde q^2 p^2x(1-x)}\Big).
\label{C.26}
\end{gather}
Expanding $\hat I^\mu(p,\tilde q)$ over the small $\tilde q$'s we have found \begin{equation}
\hat I^\mu(p,\tilde q)= \frac{1}{8\pi}\,\frac{\tilde q^\mu}{\sqrt{\tilde q^2}}+\frac{i}{16}\frac{ p^\mu}{\sqrt{ p^2}}\,+\, \hat f^\mu(p,\tilde q),
\label{C.27}
\end{equation}
where $\hat f^\mu(p,\tilde q)$ vanishes as $\tilde q^\mu\rightarrow 0$,
the expression exactly equivalent to the one in (\ref{C.21}).

\section{Feynman rules}

\subsection{Gauge fields and ghosts-gauge field vertices}

\begin{figure}[t]
\begin{center}
\includegraphics[width=10cm]{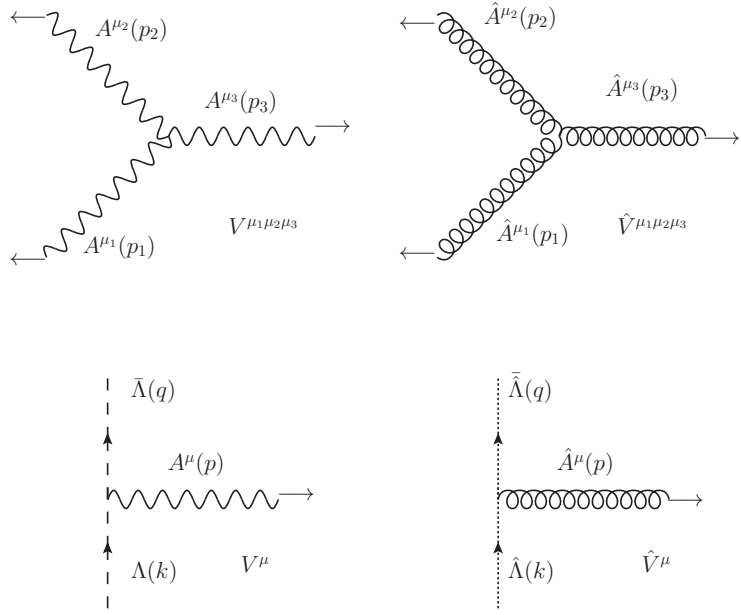}
\end{center}
\caption{Triple gauge field, -hgauge field, ghost-gauge field, and hghost-hgauge field vertices.}
\label{fig:Fig23FR}
\end{figure}

Starting with Chern-Simons action (\ref{ACS}), for triple-gauge field interaction, in accord with the first two diagrams in figure \ref{fig:Fig23FR}, we extract the following Feynman rules:
\begin{equation}
V^{\mu_1\mu_2\mu_3}=-\hat V^{\mu_1\mu_2\mu_3}
=i\frac{\kappa}{2\pi}2\sin\frac{{p_1}\theta{p_2}}{2} \epsilon^{\mu_1\mu_2\mu_3},
\label{FR3A}
\end{equation}
where we recall that
$p\theta k=p_\mu\theta^{\mu\nu}k_\nu$, and $p\theta p=p_\mu\theta^{\mu\nu}p_\nu=0$.
\begin{figure}[t]
\begin{center}
\includegraphics[width=11cm]{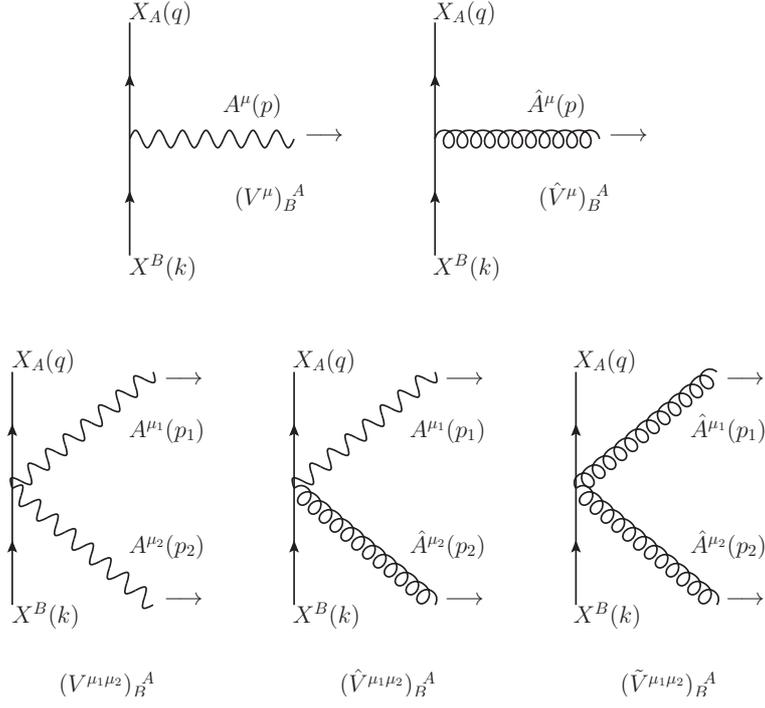}
\end{center}
\caption{Scalar-gauge field, -hgauge field vertices.}
\label{fig:Fig24FR}
\end{figure}
\begin{figure}[t]
\begin{center}
\includegraphics[width=11cm]{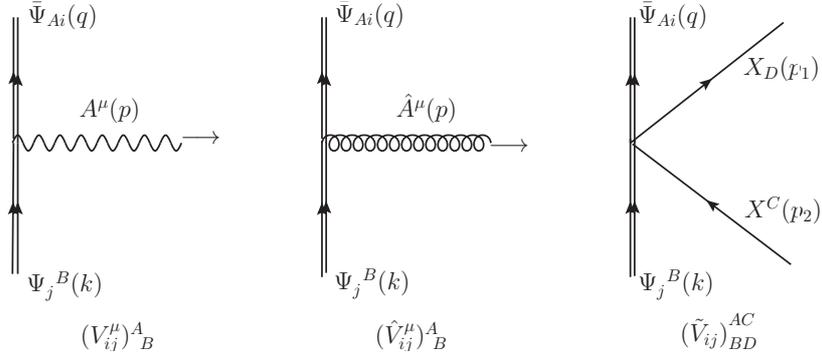}
\end{center}
\caption{Fermion-gauge field, -hgauge field and 2fermions-2scalars vertices.}
\label{fig:Fig25FR}
\end{figure}

From ghost and gauge-fixing field action (\ref{GfGh}), in accord with the second two diagrams in figure \ref{fig:Fig23FR}, we extract the following Feynman rules:
\begin{equation}
V^{\mu}=-\hat V^{\mu}=
\frac{\kappa}{2\pi}q^{\mu}\; 2\sin\frac{p\theta k}{2}.
\label{FRGfGA}
\end{equation}

\subsection{Scalar-gauge fields vertices}

From the kinetic part of the action $S_{\rm kin}$ (\ref{Akin}), in accord with figure \ref{fig:Fig24FR}, we obtain the following Feynman rules:
\begin{eqnarray}
(V^{\mu})^B_{\;\;A}&=&
i\frac{-\kappa}{2\pi}e^{\frac{i}{2}{k\theta q}}
(k+q)^\mu\delta^B_{\;\;A},
\label{FRSA}\\
(\hat V^{\mu})^B_{\;\;A}&=&
i\frac{\kappa}{2\pi}e^{-\frac{i}{2}{k\theta q}}
(k+q)^\mu\delta^B_{\;\;A},
\label{FRShA}
\end{eqnarray}
and
\begin{eqnarray}
(V^{\mu_1\mu_2})^B_{\;\;A}&=&
i\frac{-\kappa}{2\pi}\eta^{\mu_1\mu_2}
e^{\frac{i}{2}{k\theta q}}
\Big[e^{-\frac{i}{2}{p_1\theta(k-q)}} +
e^{-\frac{i}{2}{p_2\theta(k-q)}} \Big]\delta^B_{\;\;A},
\label{FR2S2hA}\\
(\hat V^{\mu_1\mu_2})^B{}_A&=&
2i\frac{\kappa}{2\pi}\eta^{\mu_1\mu_2}
\Big[e^{\frac{i}{2}{q\theta(k+p_1)}}
e^{-\frac{i}{2}{p_1\theta k}} \Big]\delta^B_{\;\;A},
\label{FR2S2A}\\
(\tilde V^{\mu_1\mu_2})^B_{\;\;A}&=&
i\frac{-\kappa}{2\pi}\eta^{\mu_1\mu_2}
e^{-\frac{i}{2}{k\theta q}}
\Big[e^{-\frac{i}{2}{p_1\theta(k-q)}} +
e^{-\frac{i}{2}{p_2\theta(k-q)}} \Big]\delta^B_{\;\;A}.
\label{FR2S1A1hA}
\end{eqnarray}

\subsection{Fermion-gauge field vertices}

From the kinetic part of the action  $S_{\rm kin}$ (\ref{Akin}), in accord with the first two terms in figure \ref{fig:Fig25FR}, we obtain relevant Feynman rules,
\begin{eqnarray}
(V^\mu_{ij})^A_{\;\;B}&=&
i\frac{-\kappa}{2\pi}\gamma^\mu_{ij}e^{-\frac{i}{2}q\theta k} \delta^A_{\;\;B},
\label{FRFA}\\
(\hat V^\mu_{ij})^A_{\;\;B}&=&
i\frac{\kappa}{2\pi}\gamma^\mu_{ij}e^{\frac{i}{2}q\theta k} \delta^A_{\;\;B},
\label{FRFhA}
\end{eqnarray}
while from the action $S_4$  (\ref{AS4}), in accord with third diagram in figure \ref{fig:Fig25FR}, we have
\begin{equation}
\Big(\tilde V_{ij}\Big)^{AC}_{BD}
=i\frac{\kappa}{2\pi}\delta_{ij}\Big(\delta^A_B\delta^C_D-2\delta^A_D\delta^C_B\Big)
2\sin\frac{q\theta k + p_1\theta p_2}{2}.
\label{FRXXPsiPsi}
\end{equation}

\end{document}